\documentclass[11pt, a4paper]{article}
\usepackage[T1]{fontenc}
\usepackage[utf8]{inputenc}
\usepackage{amsmath}
\usepackage{amssymb}
\usepackage{graphicx}
\usepackage{epstopdf}
\usepackage{epsfig}
\usepackage{fancyhdr}
\usepackage[twoside, margin=2.5cm, bindingoffset=0.0cm]{geometry}
\usepackage{url}
\usepackage{epstopdf}
\usepackage{array}
\usepackage{color}
\usepackage{easy-todo}
\usepackage{lineno,hyperref}
\usepackage{algorithm}
\usepackage{algpseudocode}
\algrenewcommand\Require{\item[\textbf{Input:}]}
\algrenewcommand\Ensure{\item[\textbf{Output:}]}

\usepackage{amsfonts}
\usepackage{tikz}
\usetikzlibrary{patterns}

\tikzset{every picture/.style={line width=0.75pt}} %set default line width to 0.75pt        
\usepackage{subcaption}
 
\usepackage{amsthm}

\usepackage{setspace}
\usepackage{easy-todo}
\usepackage{multicol}
\usepackage[backend=biber, style=numeric-comp , url=false]{biblatex}
\DeclareSourcemap{
  \maps[datatype=bibtex]{
    \map{
      \step[fieldset=issn, null]
      \step[fieldset=isbn, null]
    }
  }
}
\title{Accelerated Solvers for Neutral Particle Dynamics in Plasma Simulation}

\author{Margherita Guido\thanks{École Polytechnique Fédérale de Lausanne (EPFL), Institute of Mathematics, 1015 Lausanne, Switzerland (margherita.guido@epfl.ch, daniel.kressner@epfl.ch)}\,\,\thanks{École Polytechnique Fédérale de Lausanne (EPFL), Swiss Plasma Center (SPC), 1015 Lausanne, Switzerland (davide.mancini@epfl.ch, paolo.ricci@epfl.ch)}  \and Daniel Kressner\footnotemark[1] \and Davide Mancini\footnotemark[2] \and  Paolo Ricci\footnotemark[2]}

\begin{document}
\maketitle
%\doublespacing
%%\vspace{0pt}
%\vspace{10pt}

%\tableofcontents
\setcounter{page}{1}

\begin{abstract}
The simulation of turbulence in the boundary region of a tokamak is crucial for understanding and optimizing the performance of fusion reactors. In this work, the use of low-rank linear algebra techniques is shown to enhance the efficiency of boundary simulations, specifically by accelerating the solution of a kinetic model for the neutral particles. Solving the kinetic model deterministically using the method of characteristics requires the solution of integral equations, which typically result in dense linear systems upon discretization. We employ hierarchical matrix approximations to significantly reduce the computational cost of assembling and solving the linear systems, leading to substantial savings in both time and memory. The hierarchical matrix method is implemented and tested within the GBS simulation code for boundary simulations, achieving over 90\% reduction in computation time and memory, and enabling simulations with unprecedented spatial resolution for neutral particles.
\end{abstract}

 %\listoftodos 

\section{Introduction}
\label{sec:intro}

The physics understanding of the tokamak boundary is key in the design and operation of future nuclear fusion power plants. 
The physical phenomena occurring in this region set the boundary conditions for the core region, and control the interaction of the plasma with the external wall. 
The boundary is characterized by the presence of neutral particles, which are not affected by the magnetic field lines and interact with the plasma through collisional processes.
Indeed, as ions and electrons outflowing from the confined core region impact the solid walls, they recombine  and are reemitted as neutral atoms and molecules. Through different collisional processes, the recycled neutrals influence the dynamics of the plasma in the boundary.
Numerical simulations represent an essential tool for understanding the interaction between neutral particles and plasma.

The dynamics of neutral particles in a tokamak is historically described via kinetic Monte Carlo simulations as in the codes EIRENE~\cite{Reiter2005}, DEGAS 2~\cite{Stotler2005}, and GTNEUT \cite{Mandrekas2004}. This is often coupled to fluid codes for the simulation of plasma, where perpendicular transport is modeled as a diffusive process, as in the widely used  SOLPS-ITER~\cite{Wiesen2015} and SOLEDGE2D~\cite{Bufferand2017}. In order to simulate turbulent phenomena in the boundary,  codes based on the drift-reduced Braginskii equations~\cite{Braginskii1965} are most often used to simulate the plasma, and implemented in codes such as BOUT++~\cite{Dudson2009}, TOKAM3X~\cite{Tamain2016}, GDB~\cite{Zhu2018}, GRILLIX~\cite{Stegmeir2018}, and HESEL~\cite{Madsen2018}. The EIRENE code, based on a kinetic Monte Carlo approach, has been coupled to the turbulent plasma code TOKAM3X~\cite{Tamain2014,Baudoin2018}, resulting in SOLEDGE3X~\cite{Rivals2025}. The neutral particles code based on EIRENE has also been coupled to the fluid 3D turbulence code BOUT++ in~\cite{Zhang2019}.  In GRILLIX~\cite{Stegmeir2018}, the turbulent plasma model is coupled to a simplified diffusive neutral gas model that does not account for relevant kinetic effects.

As an alternative to the statistical Monte Carlo approach, Wersal et al.~\cite{Wersal2015} proposed a neutral particle model that solves the kinetic equations by integrating along characteristics and numerically discretizing the resulting equations.
Being deterministic, this approach overcomes the accuracy limitations inherent in Monte Carlo models. This model has been coupled to  a self-consistent evolution of the drift-reduced Braginskii equations for the plasma in the GBS code~\cite{Giacomin2022}.

%fluid approximations (diffusive or otherwise) are used for computational efficiency.

The kinetic equations integrated along the characteristics yield 
a two-dimensional linear integral equation of the form
\begin{equation}
    \label{eq:kerneleq}
    \int_{D} \mathcal{K}(x,y) f(y) \, \text{d}y = g(x) \quad \quad x \in D \subset \mathbb{R}^2,
\end{equation}
which needs to be solved numerically for $f$. The integral kernel $\mathcal{K}(x,y)$ is non-local, as it takes a non-vanishing value for each couple of optically connected points $(x,y) \in D$. After numerical discretization, the integral equation~\eqref{eq:kerneleq} becomes a linear system of equations.
When using a cartesian grid of $N$ points in the domain $D$ for this purpose, the kernel function must be evaluated $N^2$ times and a dense linear system of size $N\times N$ must be solved. This process has a high computational cost,  both in terms of memory and computing time, and represents a major limitation of the deterministic kinetic model. This is especially critical when targeting complex simulations aimed at modeling large-scale fusion reactors.

The goal of this work is to accelerate the numerical solution of the (coupled) GBS kinetic neutral model by
avoiding computations with dense matrices. This approach allows for higher-resolution neutral model simulations while maintaining a low computational cost. Note that the kernel function in~\eqref{eq:kerneleq} is not available in closed form, but it is computed from plasma quantities provided by the plasma simulation. This makes it difficult to exploit analytic techniques like the Fast Multipole Method~\cite{Greengard1997} and leads us to rely on an algebraic approach.
Specifically, we exploit a hierarchical matrix representation~\cite{Hackbusch2015,Bebendorf2008} that partitions the matrix hierarchically, identifying blocks where the kernel is smooth and thus allow for low-rank approximation. Using cross approximation, there is no need to compute all the matrix entries for such blocks. Moreover, this representation enables efficient operations such as matrix-vector multiplication, reducing the cost of iterative linear system solvers.
The compressed form results in reduced computational storage and time, from quadratic to nearly linear in $N$,  enabling the handling of large-scale problems that otherwise would not be computationally tractable.

The rest of the paper is organized as follows. In Section~\ref{sec:neutralmodel}, we describe the kinetic neutral model and its formal solution through the method of characteristics. Section~\ref{sec:numsol} is concerned with the numerical discretization of the neutral model in GBS, highlighting its most expensive parts. In Section~\ref{sec:hm}, we discuss the hierarchical approximation of the plasma-plasma interaction matrix.
In Section~\ref{sec:numres}, we demonstrate that the hierarchical matrix approximation can significantly reduce  simulation time and memory when implemented in the GBS code.

\section{Neutral Model}
\label{sec:neutralmodel}

In this section, we summarize the neutral kinetic model introduced in~\cite{Wersal2015}.
To describe the neutral particles, we consider a single mono-atomic neutral species described by a distribution function $f_{\textrm{n}}$. Its dynamics evolves according to the kinetic equation 
\begin{equation}
    \label{eq:kin}
    \frac{\partial f_{\textrm{n}}}{\partial t}  + \boldsymbol{v} \cdot \nabla f_{\textrm{n}} = - \nu_{\textrm{iz}} f_{\textrm{n}} - \nu_{\textrm{cx}} \left( f_{\textrm{n}} - \frac{n_{\textrm{n}}}{n_{\textrm{i}} f_{\textrm{i}}}\right) + \nu_{\textrm{rec}} f_{\textrm{i}},
\end{equation}
where $f_{\textrm{i}}$, $f_{\textrm{n}}$ and $n_{\textrm{i}}$, $n_{\textrm{n}}$ are the ion/neutral distribution function and density. The ionization, charge-exchange, and recombination processes, are represented in Eq.~\eqref{eq:kin} by the collision frequencies $\nu_{\textrm{iz}}, \, \nu_{\textrm{cx}}$ and $\nu_{\textrm{rec}}$, evaluated by employing the Krook operator, starting from given plasma quantities assumed to be known. 
The neutral-neutral collisions are neglected since their reaction rate is lower than ionization and charge-exchange processes, while the elastic electron-neutral collisions term is neglected in~\eqref{eq:kin} due to the electron to neutral mass ratio. 

Equation~\eqref{eq:kin} is solved in a three-dimensional domain that encompasses the whole tokamak plasma volume, modeled as a torus with arbitrary poloidal cross-section. Within the tokamak large ratio approximation, we can consider the poloidal plane $D$ perpendicular to the toroidal magnetic field $\boldsymbol{B}$. 
The equation for $f_{\textrm{n}}$ is completed by boundary conditions at the solid walls, obtained from assuming that the impacting particles (neutrals and ions) on the walls are reflected or absorbed and then reemitted. More precisely, we assume that a fraction $\alpha_{\textrm{refl}}$ of particles is reflected as neutrals, while the remaining $1- \alpha_{\textrm{refl}}$ fraction of absorbed particles is released with a velocity that depends on the wall properties and is independent of the impacting energy. 
Hence, the distribution function for the inflowing neutrals (for which $v_{p} = \boldsymbol{v}\cdot \boldsymbol{\hat{n}} > 0$, where $\boldsymbol{\hat{n}}$ is the normal vector pointing in the plasma region) is given by
\begin{align*}
    f_{\textrm{n}} (\boldsymbol{x_{b}}, \boldsymbol{v}) &= (1- \alpha_{\textrm{refl}}) \Gamma_{\textrm{out}} (\boldsymbol{x_{b}}) \chi_{\textrm{in}} (\boldsymbol{{x_b}}, \boldsymbol{v}) \nonumber \\ & +  \alpha_{\textrm{refl}} \left[  f_{\textrm{n}}  (\boldsymbol{{x_b}}, \boldsymbol{v} - 2 \boldsymbol{v_{p}}) - f_{\textrm{i}}  (\boldsymbol{{x_b}}, \boldsymbol{v} - 2 \boldsymbol{v_{p}}) \right],
\end{align*}
where $\boldsymbol{x_{b}}$ is the vector position of a point on the boundary, $\Gamma_{\textrm{out}} = \Gamma_{\textrm{out,n}}+ \Gamma_{\textrm{out,i}}$  is the sum of the fluxes of ions and neutrals to the wall, $\boldsymbol{v_{p}} = v_{p} \boldsymbol{\hat{n}}$ is the velocity perpendicular to the boundary, and $\chi_{\textrm{in}}$ is the inflowing velocity distribution. 

Assuming known  plasma properties, the kinetic equation in~\eqref{eq:kin} is formally solved with the method of characteristics, under two main assumptions, valid in the typical boundary parameters regime. 
First, it is assumed that the turbulent time-scale $\tau_{\mathrm{turb}}$ is much larger than the typical time of flight of
neutrals, $\tau_{\textrm{n}} \ll \tau_{\mathrm{turb}}$. This motivates the neutral adiabatic approximation, which assumes that the neutral distribution function is static over the turbulent timescale, $\partial_{t} f_{n } = 0$. Second, it is assumed that the neutral mean free path is shorter than typical lengths of the plasma structures along the direction parallel to the magnetic field. This leads to the reduction of the three-dimensional model to a set of  independent two-dimensional problems, one for each poloidal plane. Indeed, an arbitrary point of the domain is decomposed over two coordinates perpendicular and parallel to $\boldsymbol{B}$, $\boldsymbol{x_{\perp}} \in D$ and $x_{\|} \in T$, where $T$ denotes the toroidal direction. The coordinate $\boldsymbol{x_{\perp}}$ spans $D$, the two dimensional plane perpendicular to the magnetic field, approximately corresponding to the poloidal plane.  Moreover, we identify with $\boldsymbol{x_{\perp \textrm{b}}} \in \partial D$ a point of the boundary of the poloidal domain.

Using the above assumptions, the formal solution of Eq.~\eqref{eq:kin} using the method of characteristics becomes 
\begin{align}
    \label{eq:sol}
    f_{\textrm{n}} (\boldsymbol{x_{\perp}}, x_{\|}, \boldsymbol{v}, t) = \int_{0}^{r_{b}} &\biggl\{   \frac{S(\boldsymbol{x_{\perp}'}, x_{\|}, \boldsymbol{v}, t)  }{v_{\perp}} + \delta (r' -r_{b}) f_{\textrm{n}}  (\boldsymbol{x_{\perp}'}, x_{\|}, \boldsymbol{v}, t) \nonumber \\ & \times \exp \left[ -\frac{1}{v_{\perp}}  \int_{0}^{r_{\perp}'}    \nu_{\textrm{eff}}(\boldsymbol{x_{\perp}''}, x_{\|},  t)  \, \text{d} r''\right] \biggl \} \, \text{d} r',
\end{align}
 where the coordinate $r'$ is taken along the neutral characteristic $\boldsymbol{x_{\perp}'} = \boldsymbol{x_{\perp}} - r' \boldsymbol{v_{\perp}} / v_{\perp}$, $r'_{b}$ denotes the distance along the characteristic from the position $\boldsymbol{x}$ and the wall, $\boldsymbol{v_{\perp}}$ is the component of the velocity perpendicular to $\boldsymbol{B}$. The second, exponentially decaying term in~\eqref{eq:sol} accounts for the removal of neutrals along their trajectory due to collisions. Specifically, it involves integrating the effective collision frequency, $\nu_{\textrm{eff}}$, along the characteristic path. This effective frequency is defined as
\begin{equation} 
    \label{eq:nueff} \nu_{\textrm{eff}} = \nu_{\textrm{iz}} + \nu_{\textrm{cx}}, 
\end{equation}
which accounts for contributions from both ionization, $\nu_{\textrm{iz}}$, and charge-exchange, $\nu_{\textrm{cx}}$, processes. These interactions remove neutrals as they travel from the source location $\boldsymbol{x_{\perp}'}$ to the target location $\boldsymbol{x_{\perp}}$. 
In Eq.~\eqref{eq:sol}, single primes are used to indicate the source location of neutral particles, while double primes indicate locations along a path integral between the source  $\boldsymbol{x_{\perp}'}$ and target $\boldsymbol{x_{\perp}}$.
In the following, we drop the parametric dependence from $x_{\parallel}$, since the poloidal planes are decoupled,  and from the time $t$, since we have expressed the solution in a steady-state form.  
 The volumetric source term $S(\boldsymbol{x_{\perp}'}, \boldsymbol{v})$ in~\eqref{eq:sol} results from charge-exchange and recombination
 processes, and it depends on the neutral distribution  function through the neutral density $\boldsymbol{n}_{\textrm{n}} (\boldsymbol{x_{\perp}}) = \int f_{\textrm{n}} d\boldsymbol{v}$.
By integrating~\eqref{eq:sol} in velocity space, the following linear integral equation for the neutral density is obtained:
\begin{align}
    \label{eq:int1}
    \boldsymbol{n}_{\textrm{n}} (\boldsymbol{x_{\perp}}) = & \int_{D}  \boldsymbol{n}_{\textrm{n}} (\boldsymbol{x_{\perp}'}) \nu_{\textrm{cx}} (\boldsymbol{x_{\perp}'}) K_{p \rightarrow p }(\boldsymbol{x_{\perp }}, \boldsymbol{x_{\perp}'}) \, \text{d}A' \nonumber \\
    & + \int_{\partial D} (1-\alpha_{\textrm{refl}}) \Gamma_{\textrm{out,n} }(\boldsymbol{x_{\perp b}'}) K_{b \rightarrow p }(\boldsymbol{x_{\perp }}, \boldsymbol{x_{\perp b}'}, T_{b})\, \text{d}a_{b}' \nonumber \\
    & + n_{n\left[\textrm{out,i}\right]} (\boldsymbol{x_{\perp }}) +n_{n\left[\textrm{rec}\right]} (\boldsymbol{x_{\perp }})m
\end{align}
where $\text{d}A'$ is the infinitesimal area in the poloidal plane $D$, $\text{d}a'_{b}$ is the infinitesimal length along the boundary $\partial D$, and $\Gamma_{\textrm{out,n} }(\boldsymbol{x_{\perp b}})$ is the neutral flux towards the wall:
\begin{align}
    \label{eq:int2}
    \Gamma_{\textrm{out,n} }(\boldsymbol{x_{\perp b}})= & \int_{D}  \boldsymbol{n}_{\textrm{n}} (\boldsymbol{x_{\perp}'}) \nu_{\textrm{cx}} (\boldsymbol{x_{\perp}'}) K_{p \rightarrow b }(\boldsymbol{x_{\perp b}}, \boldsymbol{x_{\perp}'}) \, \text{d}A' \nonumber \\
    & + \int_{\partial D} (1-\alpha_{\textrm{refl}}) \Gamma_{\textrm{out,n} }(\boldsymbol{x_{\perp b}'}) K_{b \rightarrow b }(\boldsymbol{x_{\perp b}}, \boldsymbol{x_{\perp b}'}, T_{b}) \, \text{d}a_{b}' \nonumber \\
    & +  \Gamma_{\textrm{out,n} \left[\textrm{out,i}\right]} (\boldsymbol{x_{\perp b}}) + \Gamma_{\textrm{out,n}\left[\textrm{rec}\right]} (\boldsymbol{x_{\perp b}}) 
\end{align}
The integrals in equations~\eqref{eq:int1} and~\eqref{eq:int2} are computed using cylindrical coordinates in velocity space: $(v_{\perp}, \theta, v_{\|})$.

The two source terms in equations~\eqref{eq:int1} and~\eqref{eq:int2} represent, respectively, the contribution to the neutral density and flux due to the ion recycling at the wall and recombination events:
\begin{align*}
%  \label{eq:sources_n}
  n_{n\left[\textrm{out,i}\right]} (\boldsymbol{x_{\perp }}, T_{b} ) =  & \int_{\partial D}
  \Gamma_{\textrm{out,i}}(\boldsymbol{x_{\perp b}'})
   \left[
  (1-\alpha_{\textrm{refl}} ) K_{b \rightarrow p }(\boldsymbol{x_{\perp}}, \boldsymbol{x_{\perp b}'}, T_{b} ) \right. \nonumber
 \\  
  & + \left.\alpha_{\textrm{refl}} K_{b \rightarrow p }
  (\boldsymbol{x_{\perp}}, \boldsymbol{x_{\perp b}'}, T_{\textrm{i}})
   \right] \, \text{d}a_{b}'\,,  \\
  n_{n\left[\textrm{rec}\right]} (\boldsymbol{x_{\perp }}) =  & \int_{ D}
   n_{\textrm{i}}(\boldsymbol{x_{\perp}'}) \nu_{\textrm{rec}}(\boldsymbol{x_{\perp}'})    K_{p \rightarrow p }(\boldsymbol{x_{\perp}}, \boldsymbol{x_{\perp }'} ) \, \text{d}A'\,,  \\
%  \label{eq:sources_gamma}
  \Gamma_{\textrm{out,n}\left[\textrm{out,i}\right]} (\boldsymbol{x_{\perp b}}, T_{b} ) =  & \int_{\partial D}
  \Gamma_{\textrm{out,i}}(\boldsymbol{x_{\perp b}'})
   \left[
  (1-\alpha_{\textrm{refl}} ) K_{b \rightarrow b }(\boldsymbol{x_{\perp b }}, \boldsymbol{x_{\perp b}'}, T_{b} ) \right. \nonumber
 \\  
  & + \left.\alpha_{\textrm{refl}} K_{b \rightarrow b }
  (\boldsymbol{x_{\perp b}}, \boldsymbol{x_{\perp b}'}, T_{\textrm{i}})
   \right] \, \text{d}a_{b}' \,, \\
 \textrm{and }  \,  \Gamma_{\textrm{out,n}\left[\textrm{rec}\right]} (\boldsymbol{x_{\perp b}}) =  & \int_{ D}
   n_{\textrm{i}}(\boldsymbol{x_{\perp}'}) \nu_{\textrm{rec}}(\boldsymbol{x_{\perp}'})    K_{p \rightarrow b}(\boldsymbol{x_{\perp b}}, \boldsymbol{x_{\perp }'} ) \, \text{d}A'.
\end{align*}

The kernel functions $K_{p \rightarrow p }, K_{p \rightarrow b }, K_{b \rightarrow p }$ and $K_{b \rightarrow b }$, in Eqs.~\eqref{eq:int1} and~\eqref{eq:int2}, represent the four probabilities for a neutral particle, generated in the plasma ($p$) or at the boundary ($b$) at the points $\boldsymbol{x_{\perp }'}$ or $\boldsymbol{x_{\perp b }'}$, of ending up in  $\boldsymbol{x_{\perp }}$ or $\boldsymbol{x_{\perp b }}$ depending on whether it is in the plasma ($p$) or the boundary ($b$). These kernels consider the direct path between two points, as well as paths that include one reflection at the wall. 
In particular, the kernel $K_{p \rightarrow p }(\boldsymbol{x_{\perp }}, \boldsymbol{x_{\perp}'}) $ takes the form
\begin{equation*}
%  \label{eq:Kpp}
     K_{p \rightarrow p }(\boldsymbol{x_{\perp }}, \boldsymbol{x_{\perp}'}) = K^{\textrm{dir}}_{p \rightarrow p }(\boldsymbol{x_{\perp }}, \boldsymbol{x_{\perp}'})  + \alpha_{\textrm{refl}} K^{\textrm{refl}}_{p \rightarrow p }(\boldsymbol{x_{\perp }}, \boldsymbol{x_{\perp}'})  \quad \quad
\end{equation*}
for all  possible paths that start and end in the plasma region, that is, for all $\boldsymbol{x_{\perp }}, \boldsymbol{x_{\perp}'} \in D \subset \mathbb{R}^2 $.
For $\textrm{path} = \left \{ \textrm{refl}, \textrm{dir}\right \}$, we  define 
\begin{equation*}
    K^{\textrm{path}} _{p \rightarrow p }(\boldsymbol{x_{\perp }}, \boldsymbol{x_{\perp}'})= \frac{1}{r'}  \int_{0}^{\infty} \Phi_{\perp\textrm{i}} (\boldsymbol{x_{\perp }'}, \boldsymbol{v_{\perp}}) \, \text{exp}\left[ -\frac{1}{v_{\perp}} \int_{0}^{r' } \nu_{\textrm{eff}} ( \boldsymbol{x_{\perp }''} )\, \text{d} r'' \right] \text{d}v_{\perp},
\end{equation*}
where $\Phi_{\perp\textrm{i}}$ is the ion velocity distribution integrated along the parallel component, and the vector $\boldsymbol{x_{\perp }''}$ represents the position along the path from the source to the target points. 
More precisely, the exponentially decaying term takes into account the loss of neutrals between the origin and arrival positions due to ionization and charge-exchange collisions. The integral $\int_{0}^{r' } \nu_{\textrm{eff}} ( \boldsymbol{x_{\perp }''} )\, \text{d} r'' $  is computed along the line that connects  $\boldsymbol{x_{\perp }'}$ and  $\boldsymbol{x_{\perp }}$, respectively for $\textrm{path} = \left \{ \textrm{refl}, \textrm{dir}\right \}$, a direct line or a path of two line segments with a reflection at one wall. The effective collision frequency $\nu_{\textrm{eff}}$ as defined in~\eqref{eq:nueff} does not depend on $f_{\textrm{n}}$ or any of its moments, and it can be evaluated from the plasma quantities, as it will be described in Section~\ref{sec:numsol}. Given that $r'$ is the euclidean norm of the distance between the two considered points, i.e. $r'=\| \boldsymbol{x_{\perp}} - \boldsymbol{x_{\perp}'} \|_{2}$, the kernels have a weak singularity when $\boldsymbol{x_{\perp}} = \boldsymbol{x_{\perp}'}$, that is, when we are trying to compute a possible path that starts and arrives in the same point.
The kernels $K_{p \rightarrow b }, K_{b \rightarrow p }$ and $K_{b \rightarrow b }$ have similar expressions that can be found in~\cite{Wersal2017}.

To summarize, equations~\eqref{eq:int1} and~\eqref{eq:int2} are two-dimensional linear integral equations involving  non-local, data-based kernel operators. Once equations~\eqref{eq:int1} and~\eqref{eq:int2} are solved, it is possible to compute higher moments of the distribution function $f_{\textrm{n}}$ via~\eqref{eq:sol}, such the neutral velocity, temperature and flux.

The model in equations~\eqref{eq:int1} and~\eqref{eq:int2} has been implemented in  GBS (Global Braginskii Solver), a first-principles, three-dimensional, flux-driven, global, turbulence code that evolves the Braginskii equations in the drift-reduced approximation to simulate the dynamics of plasma. The plasma drift-reduced equations depend on the kinetic neutral model through the neutral density, temperature and velocity, used to evaluate source or sink terms representing ionization and recombination processes~\cite{Giacomin2022}. On the other hand, plasma quantities such as the electron and ion temperature, density and ion flux as well as the one of the reaction rates, are used in the neutral kinetic equations to compute the collision frequencies in Eq.~\eqref{eq:kin}, generating a full coupling of the two models. In our description of the solution of the neutral model, we consider the plasma quantities known, assuming that they are computed as output of a generic fluid model. 
We finally note that the model for neutral particles presented here was extended in~\cite{Coroado2021} to a multispecies neutral model.

\section{Numerical Discretization of Neutral Model}

\label{sec:numsol}
\begin{figure}[H]
    %\centering
    \begin{subfigure}[b]{0.45\textwidth}
        \centering
        \includegraphics[width=\textwidth]{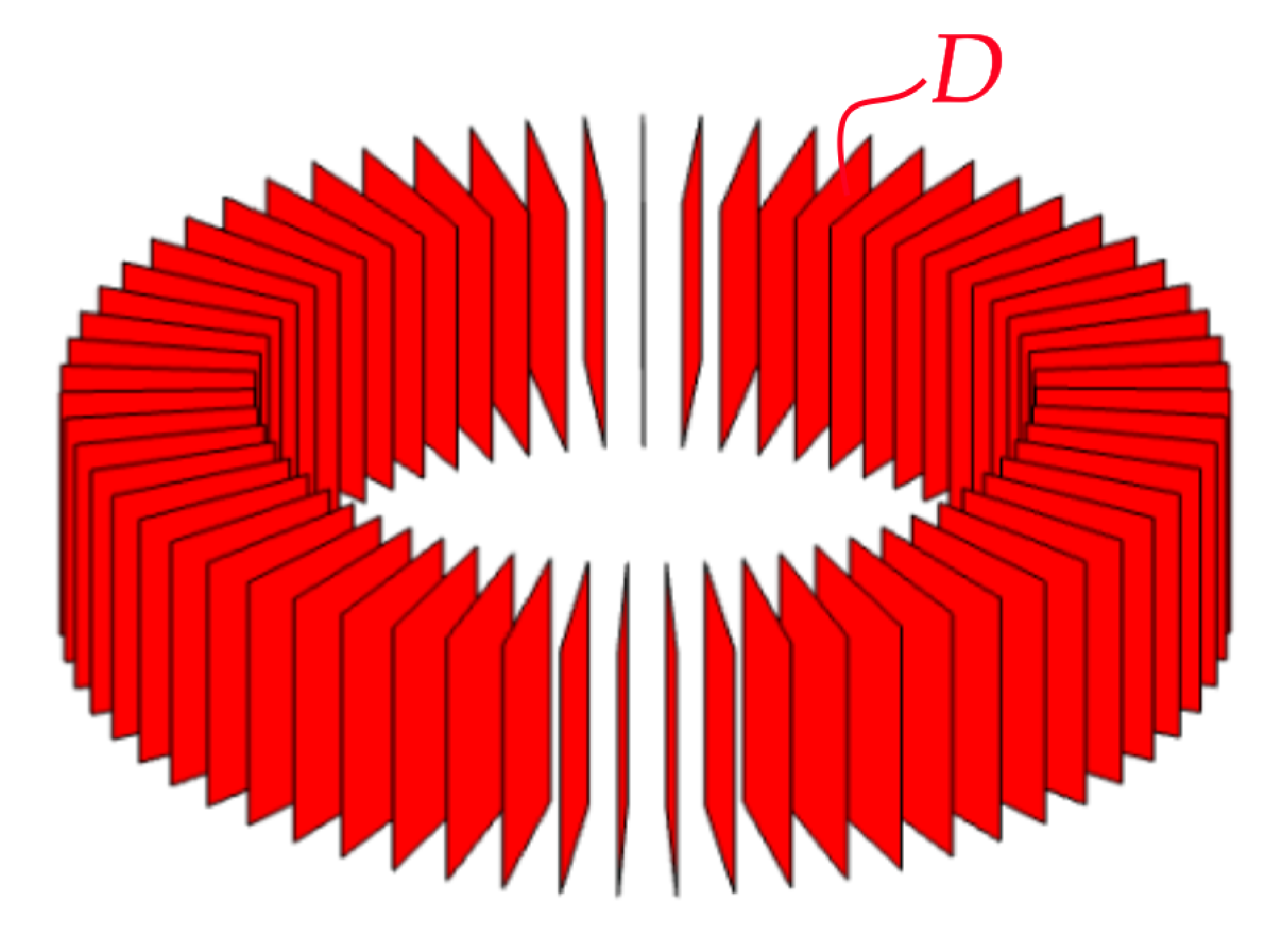} 
        \vspace{5mm}
        \subcaption{ }
        \label{fig:1_1}
    \end{subfigure}
    % \hfill
    \hspace{16mm}
    \begin{subfigure}[b]{0.4\textwidth}
        \centering
        \input{TikzFigure1}
        \subcaption{ }
        \label{fig:1_2}
    \end{subfigure} 
    \caption{\centering Discretization of the computational domain,  along the toroidal direction in $N_{\parallel}$ poloidal planes (a) and of one poloidal plane $D$ in $N_{x} \times N_{y}$ points (b). }
    \label{fig:1}
\end{figure}

We aim at numerically computing  the neutral density and flux in the domain $(\boldsymbol{x_{\perp}}, x_{\parallel}) \in D \times T$, by solving equations~\eqref{eq:int1} and~\eqref{eq:int2} in the poloidal domain $D \subset \mathbb{R}^2$ for fixed $ x_{\parallel} \in  T \subset \mathbb{R}$. For this purpose, we use a cartesian spatial  discretization. We consider $N_{\parallel}$ points to discretize the toroidal direction $x_{\parallel} \in T$, generating $N_{\parallel}$ poloidal planes, as shown in Figure~\ref{fig:1_1}. For each poloidal plane, $ \boldsymbol{x_{\perp}} \in D$	is discretized on an $N_{x} \times N_{y}$ cartesian grid, as shown  in Figure~\ref{fig:1_2}. The boundary region  of one poloidal plane is indicated in green and it consists of $N_B = 2(N_{x} + N_{y})$ points, while the internal plasma region indicated in blue consists of $N_P = N_{x}N_{y}$ points. We indicate with $N= N_P + N_B$ the total number of points involved in the discretization of $D$.
At some fixed time $t$ and toroidal coordinate $x_\parallel$, our aim is to approximate the discretized neutral density $\hat{\boldsymbol{n}}_{\textrm{n}}  \in \mathbb{R}^{N_P}$ and density flux $\hat{\boldsymbol{\Gamma}}_{\textrm{out,n}} \in \mathbb{R}^{N_B}$ on the poloidal cartesian grid.

With the spatial discretization described above, we use the collocation method~\cite{Atkinson1997} to discretize the  integral equations~\eqref{eq:int1} and~\eqref{eq:int2}, resulting in a linear system of the form 
\begin{equation}
  \label{eq:linsys}
  \Bigg(I-
  \underbrace{ \begin{bmatrix}
    \nu_{\textrm{cx}}   \hat{K}_{p \rightarrow p} &  (1 - \alpha_{\textrm{refl}}
    ) \hat{K}_{b \rightarrow p}( T_{b} )\\
     \nu_{\textrm{cx}} \hat{K}_{p \rightarrow b} &  (1 - \alpha_{\textrm{refl}}) \hat{K}_{b \rightarrow b}( T_{b} )
    \end{bmatrix}}_{K}  \Bigg) \begin{bmatrix}
    \hat{\boldsymbol{n}}_{\textrm{n}} \\ \hat{\boldsymbol{\Gamma}}_{\textrm{out,n}} \end{bmatrix}= \begin{bmatrix}
         \hat{\boldsymbol{n}}_{n\left[\textrm{out,i}\right]} +\hat{\boldsymbol{n}}_{n\left[\textrm{rec}\right]}  \\
          \hat{\boldsymbol{\Gamma}}_{\textrm{out,n} \left[\textrm{out,i}\right]}  + \hat{\boldsymbol{\Gamma}}_{\textrm{out,n}\left[\textrm{rec}\right]}
  \end{bmatrix}.
\end{equation}
The terms on the right-hand side terms are computed as
\begin{align*}
%  \label{eq:sources_n_disc}
  &\hat{\boldsymbol{n}}_{n\left[\textrm{out,i}\right]}   =  \left[(1 - \alpha_{\textrm{refl}}) \hat{K}_{b \rightarrow p}( T_{b} ) + \alpha_{\textrm{refl}} \hat{K}_{b \rightarrow p}( T_{\textrm{i}} )\right] \hat{\boldsymbol{\Gamma}}_{\textrm{out,i}} 
  \\  
  &\hat{\boldsymbol{n}}_{n\left[\textrm{rec}\right]}  = \nu_{\textrm{rec}}\hat{K}_{p \rightarrow p} \hat{\boldsymbol{n}}_{\textrm{i}} \\
%  \label{eq:sources_gamma_disc}
  &\hat{\boldsymbol{\Gamma}}_{\textrm{out,n} \left[\textrm{out,i}\right]}  =  \left[(1 - \alpha_{\textrm{refl}}) \hat{K}_{b \rightarrow b}( T_{b} ) + \alpha_{\textrm{refl}} \hat{K}_{b \rightarrow b}( T_{\textrm{i}} )\right] \hat{\boldsymbol{\Gamma}}_{\textrm{out,i}} 
 \\  
 &\hat{\boldsymbol{\Gamma}}_{\textrm{out,n}\left[\textrm{rec}\right]} = \nu_{\textrm{rec}}\hat{K}_{p \rightarrow b} \hat{\boldsymbol{n}}_{\textrm{i}}.
\end{align*}
The four submatrices of the matrix $K  \in \mathbb{R}^{N \times N}$ in Eq.~\eqref{eq:linsys} are discretized 2-dimensional non-local integral operators, each corresponding to one of the four kernels introduced in Section~\ref{sec:neutralmodel}.
Since they exhibit  similar properties, we will hereafter focus on the first submatrix $\hat{K}_{p \rightarrow p}$, which corresponds to plasma to plasma interaction.  This submatrix is the discretization of the integral operator with kernel function $ K_{p \rightarrow p }(\boldsymbol{x_{\perp }}, \boldsymbol{x_{\perp}'}) = K^{\textrm{dir}}_{p \rightarrow p }(\boldsymbol{x_{\perp }}, \boldsymbol{x_{\perp}'})  + \alpha_{\textrm{refl}} K^{\textrm{refl}}_{p \rightarrow p }\left(\boldsymbol{x_{\perp }}, \boldsymbol{x_{\perp}'}\right)$ and its entries satisfy
\begin{equation}
    \label{eq:discKpp}
  (\hat{K}_{p \rightarrow p})_{ij} \approx \Delta A_j K_{p \rightarrow p}(\boldsymbol{x}_{\perp }^{i}, \boldsymbol{x}_{\perp}^{j})  =  \Delta A_j( K^{\textrm{dir}}_{p \rightarrow p }(\boldsymbol{x}_{\perp }^{i}, \boldsymbol{x}_{\perp}^{j})  + \alpha_{\textrm{refl}} K^{\textrm{refl}}_{p \rightarrow p }(\boldsymbol{x}_{\perp }^{i}, \boldsymbol{x}_{\perp}^{j}) ),
\end{equation}
where $\Delta A_j$ is the area of the grid square centered around $\boldsymbol{x}_{\perp}^j$. Each entry of $\hat{K}_{p \rightarrow p}$ represents the loss or gain of neutral particles along a path connecting two points in the discretized interior region of the two-dimensional poloidal domain $D$, as shown in Figure~\ref{fig:1_2}.

\textbf{Numerical Quadrature}. To compute an entry of $\hat{K}_{p \rightarrow p}$, we need to evaluate the kernel $K^{\textrm{path}} _{p \rightarrow p }$ for $\textrm{path} = \textrm{dir}$ and $\textrm{path} = \textrm{refl}$. This requires the numerical computation of integrals of the form
\begin{equation}
    K^{\textrm{path}} _{p \rightarrow p } (\boldsymbol{x}_{\perp }^{i}, \boldsymbol{x}_{\perp}^{j})  =  \frac{1}{r'} \int_{0}^{\infty} \Phi_{ i} (\boldsymbol{x_{\perp }^{j}}, \boldsymbol{v_{\perp}}) \, \text{exp}\left[ -\frac{1}{v_{\perp}} \int_{0}^{r' } \nu_{\textrm{eff}} ( \boldsymbol{x_{\perp }''} )\, \text{d} r'' \right] \text{d}v_{\perp},
  \quad \forall \,i,j= 1, \cdots , N_P \nonumber.
\end{equation}
Given the Maxwellian ion velocity distribution 
\begin{equation*}
    \Phi_{ i} (\boldsymbol{x_{\perp }^{j}}, \boldsymbol{v_{\perp}}) =\big[ m_i /(2 \pi T_i(x_{\perp }^{j} )\big]^{3/2} \exp\big(-m_i \boldsymbol{v_{\perp}}^2 /(2 T_i(x_{\perp }^{j}) ) \big),
\end{equation*} the outer integral (in velocity space) is computed using Gauss-Laguerre quadrature to account for the infinite integration domain. The inner integral $\int_{0}^{r' } \nu_{\textrm{eff}} ( \boldsymbol{x_{\perp }''} )\, \text{d} r''$ is computed with the composite trapezoidal rule along the coordinate that spans the line connecting $\boldsymbol{x}_{\perp }^{i}$ and $\boldsymbol{x}_{\perp }^{j}$, either with a direct line for the direct path or with two segments for particles reflected once at the wall. The discretized effective collision frequency
$\nu_{\textrm{eff}}$ is obtained as an output from the plasma model, on a grid that is independent of the neutral particles grid and that typically has finer resolution. Its values at the quadrature nodes are obtained via piecewise linear interpolation.
To regularize the weak singularity in the points $ \boldsymbol{x}_{\perp }^{i}=\boldsymbol{x}_{\perp}^{j}$, we set $r'=\sqrt{d}$ when $i = j$, where $d$ is the diagonal length of the cell centered in $ \boldsymbol{x}_{\perp }^{i}$.

\textbf{Properties of the linear system and numerical challenges.} The matrix $K$ defined in Eq.~\eqref{eq:linsys}, obtained after discretizing the integral operators, is densely populated due to the non-local nature of the kernel: The kernel is nonzero for every pair of points in the domain. As a consequence, we found that in a typical simulation more than 90\% of the total computational time is spent on assembling the matrix. 
Moreover, the linear system~\eqref{eq:linsys} must be independently constructed and solved for each of the $N_{\|}$ poloidal planes and additional computational effort is required for computing higher moments of the distribution function, requiring the assembly of $3 \times N_{\parallel}$ additional dense matrices with the same structure as $K$~\cite{Coroado2021}. This make the computational cost of solving equation~\eqref{eq:linsys} prohibitive at grid size of interest for large fusion devices as ITER, not only in terms of time but also due to the memory required to store the dense matrices. 

Let us stress that both the matrix and the right-hand side of the linear system in~\eqref{eq:linsys} depend on time. Since $\nu_{\textrm{eff}}$ varies in time, the kernel needs to be updated periodically and, hence, the  linear system that discretizes the integral operator needs to be reassembled.

Given a cartesian grid of size $N_{x} \times N_{y}$ and recalling that $N_B = 2(N_{x} + N_{y})$, $N_P =N_{x} N_{y}$, the submatrix $\hat{K}_{p \rightarrow p}$ contains $N_P^2$  entries. This is significantly larger than the total number of entries in the three other submatrices, given by $N_B^2 + 2 (N_P+ N_B) $.
As a result, computing and storing the entries of $\hat{K}_{p \rightarrow p}$  represents the primary computational bottleneck for performing large-scale simulations.

Despite the presence of exponentially decaying terms in the integral kernels, the resulting decay of the matrix entries away from the diagonal is too modest to render thresholding techniques effective at reducing computational cost.

\section{Acceleration via Hierarchical Matrix Approximation}
\label{sec:hm}

As discussed on Section~\ref{sec:numsol}, the assembly and storage of the discretized integral operator $\hat{K}_{p \rightarrow p}$ represents a severe computational bottleneck when solving the neutral model.
The underlying integral kernel~\eqref{eq:discKpp} is data-based, depending on the effective collision frequency $\nu_{\textrm{eff}}$ returned from the plasma model. This dependence makes it difficult to apply techniques, such as the Fast Multipole Method~\cite{Greengard1997}, that require explicit access to the kernel function. Consequently, we adopt an algebraic approach based on hierarchical matrices~\cite{Hackbusch2015}, a technique that exploits the low-rank structure of certain submatrices within a dense matrix to achieve efficient storage and computation. 

Once the matrix is set up, the linear system~\eqref{eq:linsys} is solved numerically using GMRES (Generalized Minimal Residual Method~\cite{Golub2013}). Through numerical experiments, we have verified that no preconditioning is required to achieve reasonably fast convergence. For example, for a matrix $K$ of size $N \times N = 5300 \times 5300$, GMRES applied to $I-K$ requires $10$ iterations to converge below a tolerance of $10^{-10}$. As a consequence, it suffices to realize (approximate) matrix-vector products with the submatrix $\hat{K}_{p \rightarrow p}$, in addition to assembling and storing this matrix.

\textbf{Low-rank Matrices.} Let us recall that an $m \times n$ matrix $A$ rank $r$ admits a factorization of the form \begin{equation}
    A = UV^{T}\,  ,\quad U  \in \mathbb{R}^{m \times r}, \quad V  \in \mathbb{R}^{n \times r}. \nonumber
\end{equation} 
In particular, for a low-rank matrix ($r \ll m,n$), the cost of storing $A$ reduces significantly 
We define $A$ as low-rank if $\text{rank}(A) \ll m,n $. For low-rank matrices the cost of storing $A$ and multiplying it with a vector reduces significantly from $\mathcal O(mn)$ to $\mathcal O( {r(n + m)} )$ when working with the factors $U,V$ instead of $A$.

Letting $\sigma_1 \ge \sigma_2 \ge \cdots \ge 0$ denote the singular values of $A$, the best rank-$r$ approximation error for $A$ is bounded in the 2-norm by $\sigma_{r+1}$.
Thus, while matrices rarely have low rank, they can often be well approximated by a low-rank matrix, provided that their singular values decay sufficiently fast. In particular, if $A$ arises from the discretization of an integral operator with a \emph{smooth} kernel function, it usually features fast singular value decay and, hence, admits a good low-rank approximation~\cite{Sauter2010}. 

Based on the previous discussion, one cannot expect that the matrix $\hat{K}_{p \rightarrow p}$ as a whole admits a good low-rank approximation, due the non-smoothness of the underlying kernel at the singularity $\{ x_{\perp}  = x_{\perp}' \}$. On the other hand, one can expect that submatrices of $\hat{K}_{p \rightarrow p}$ associated with subdomains not affected by this singularity can be well approximated.

\begin{figure}[H]
    \centering
    \begin{subfigure}[t]{0.49\textwidth}
        \centering
        \scalebox{0.52}{

\begin{tikzpicture}[x=0.75pt,y=0.75pt,yscale=-1,xscale=1]
%uncomment if require: \path (0,347); %set diagram left start at 0, and has height of 347

%Straight Lines [id:da839175791727367] 
\draw [color={rgb, 255:red, 123; green, 179; blue, 245 }  ,draw opacity=1 ][line width=1.5]    (309.41,20.69) -- (309.41,293.16) ;
%Straight Lines [id:da605327693435569] 
\draw [color={rgb, 255:red, 123; green, 179; blue, 245 }  ,draw opacity=1 ][line width=1.5]    (243.02,225.04) -- (508.56,225.04) ;
%Straight Lines [id:da02688421740383351] 
\draw [color={rgb, 255:red, 123; green, 179; blue, 245 }  ,draw opacity=1 ][line width=1.5]    (375.79,20.69) -- (375.79,293.16) ;
%Straight Lines [id:da6474129284471886] 
\draw [color={rgb, 255:red, 123; green, 179; blue, 245 }  ,draw opacity=1 ][line width=1.5]    (243.02,156.92) -- (508.56,156.92) ;
%Straight Lines [id:da8634620857739133] 
\draw [color={rgb, 255:red, 123; green, 179; blue, 245 }  ,draw opacity=1 ][line width=1.5]    (243.02,88.8) -- (508.56,88.8) ;
%Straight Lines [id:da44928734060261566] 
\draw [color={rgb, 255:red, 123; green, 179; blue, 245 }  ,draw opacity=1 ][line width=1.5]    (442.17,20.69) -- (442.17,262.79) -- (442.17,293.16) ;
%Shape: Rectangle [id:dp6380965543337371] 
\draw  [line width=1.5]  (243.02,20.69) -- (508.56,20.69) -- (508.56,293.16) -- (243.02,293.16) -- cycle ;
%Straight Lines [id:da7265546686643182] 
\draw [color={rgb, 255:red, 123; green, 179; blue, 245 }  ,draw opacity=1 ][fill={rgb, 255:red, 255; green, 255; blue, 255 }  ,fill opacity=1 ][line width=2.25]    (41.36,156.92) -- (175.81,156.92) ;
%Straight Lines [id:da1252098081296359] 
\draw [color={rgb, 255:red, 123; green, 179; blue, 245 }  ,draw opacity=1 ][fill={rgb, 255:red, 255; green, 255; blue, 255 }  ,fill opacity=1 ][line width=2.25]    (108.58,20.69) -- (108.58,293.16) ;
%Shape: Rectangle [id:dp07074721184796984] 
\draw  [line width=2.25]  (41.36,20.69) -- (175.81,20.69) -- (175.81,293.16) -- (41.36,293.16) -- cycle ;
%Straight Lines [id:da05412941516561931] 
\draw    (479,123.34) -- (520.65,90.21) ;
\draw [shift={(523,88.34)}, rotate = 141.5] [fill={rgb, 255:red, 0; green, 0; blue, 0 }  ][line width=0.08]  [draw opacity=0] (10.72,-5.15) -- (0,0) -- (10.72,5.15) -- (7.12,0) -- cycle    ;
%Shape: Rectangle [id:dp869649177833316] 
\draw  [color={rgb, 255:red, 0; green, 0; blue, 0 }  ,draw opacity=1 ] (540.66,127.07) -- (590.82,127.07) -- (590.82,177.23) -- (540.66,177.23) -- cycle ;

% Text Node
\draw (527.15,75.03) node [anchor=north west][inner sep=0.75pt]  [font=\large] [align=left] {{\large \textbf{Dense}}};
% Text Node
\draw (515.62,137.53) node [anchor=north west][inner sep=0.75pt]  [font=\large]  {$m$};
% Text Node
\draw (556.74,104.43) node [anchor=north west][inner sep=0.75pt]  [font=\large]  {$n$};
% Text Node
\draw (154,301.4) node [anchor=north west][inner sep=0.75pt]  [font=\LARGE]  {$D$};
% Text Node
\draw (215.75,32.45) node [anchor=north west][inner sep=0.75pt]  [font=\Huge,color={rgb, 255:red, 123; green, 179; blue, 245 }  ,opacity=1 ] [align=left] {1};
% Text Node
\draw (215.75,97.16) node [anchor=north west][inner sep=0.75pt]  [font=\Huge,color={rgb, 255:red, 123; green, 179; blue, 245 }  ,opacity=1 ] [align=left] {2};
% Text Node
\draw (215.75,165.28) node [anchor=north west][inner sep=0.75pt]  [font=\Huge,color={rgb, 255:red, 123; green, 179; blue, 245 }  ,opacity=1 ] [align=left] {3};
% Text Node
\draw (215.75,236.8) node [anchor=north west][inner sep=0.75pt]  [font=\Huge,color={rgb, 255:red, 123; green, 179; blue, 245 }  ,opacity=1 ] [align=left] {4};
% Text Node
\draw (255.85,300) node [anchor=north west][inner sep=0.75pt]  [font=\Huge,color={rgb, 255:red, 123; green, 179; blue, 245 }  ,opacity=1 ] [align=left] {1};
% Text Node
\draw (326.38,300) node [anchor=north west][inner sep=0.75pt]  [font=\Huge,color={rgb, 255:red, 123; green, 179; blue, 245 }  ,opacity=1 ] [align=left] {2};
% Text Node
\draw (396.92,300) node [anchor=north west][inner sep=0.75pt]  [font=\Huge,color={rgb, 255:red, 123; green, 179; blue, 245 }  ,opacity=1 ] [align=left] {3};
% Text Node
\draw (467.45,300) node [anchor=north west][inner sep=0.75pt]  [font=\Huge,color={rgb, 255:red, 123; green, 179; blue, 245 }  ,opacity=1 ] [align=left] {4};
% Text Node
\draw (514.33,291) node [anchor=north west][inner sep=0.75pt]  [font=\LARGE] [align=left] {$\displaystyle \hat{K}_{p\rightarrow p}$};
% Text Node
\draw (61.79,201.31) node [anchor=north west][inner sep=0.75pt]  [font=\Huge,color={rgb, 255:red, 123; green, 179; blue, 245 }  ,opacity=1 ] [align=left] {1};
% Text Node
\draw (129.01,201.31) node [anchor=north west][inner sep=0.75pt]  [font=\Huge,color={rgb, 255:red, 123; green, 179; blue, 245 }  ,opacity=1 ] [align=left] {2};
% Text Node
\draw (132.38,71.94) node [anchor=north west][inner sep=0.75pt]  [font=\Huge,color={rgb, 255:red, 123; green, 179; blue, 245 }  ,opacity=1 ] [align=left] {3};
% Text Node
\draw (64.43,71.94) node [anchor=north west][inner sep=0.75pt]  [font=\Huge,color={rgb, 255:red, 123; green, 179; blue, 245 }  ,opacity=1 ] [align=left] {4};

\end{tikzpicture}
}
        \subcaption{Level 1}
        \label{fig:2_1}
    \end{subfigure}
    %\hfill
    \begin{subfigure}[t]{0.49\textwidth}
      \centering
      \input{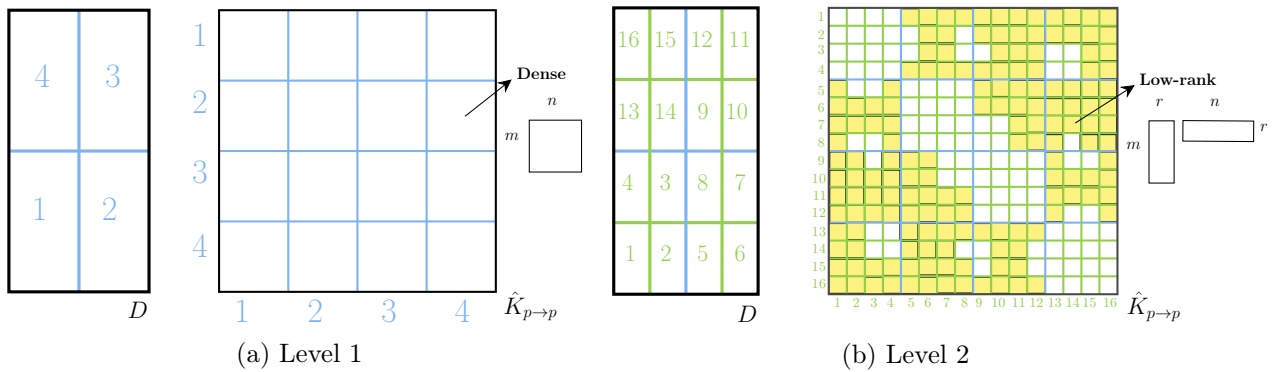}
      \vspace{-4.6mm}
      \subcaption{Level 2}
      \label{fig:2_2}
    \end{subfigure} 
    \centering
    \caption{Partitioning of the physical domain $D$ and corresponding subdivision of the matrix $\hat{K}_{p \rightarrow p}$, for (a) level 1 and (b) level 2 of the cluster tree. The admissible and inadmissible blocks are identified using criterion~\eqref{eq:admiss} and they are depicted in yellow and white, respectively. }
    \label{fig:2}
\end{figure}

\textbf{Hierarchical Matrix Representation.} Motivated by the observation made above, hierarchical matrices~\cite{Hackbusch2015} are a systematic framework for representing
and compressing matrices with low-rank blocks. In order to organize the grid points and identify matrix blocks that can be well approximated, we first partition the computational domain using a quadtree data structure. On the coarsest level, the domain $D$ is divided into four subdomains, each containing a large subset of the grid points.  On each subsequent level, each subdomain is further subdivided into four smaller subdomains, creating finer partitions. Such a domain partitioning is efficiently represented by a so called cluster tree, where the root represents the whole domain $D$ and each level of the tree corresponds to a partition of the entire domain $D$. Each node is a subdomain $C_i  \subset D$ and its four children correspond to the four subdomains of $C_i$. The recursive partitioning is stopped until the subdomains at the leaves contain a sufficiently small number of grid points.  The left side of Figures~\ref{fig:2_1} and~\ref{fig:2_2} show the partitioning of $D$ at levels 1 and 2 of the cluster tree. The Cartesian product of a partitioning of $D$ with itself induces a partitioning of the domain $D \times D$ of the integral operator and, thus, a subdivision of its discretization $\hat K_{p\to p}$, as shown on the right side of Figures~\ref{fig:2_1} and~\ref{fig:2_2}.

 \begin{figure}[H]
    \centering
    \input{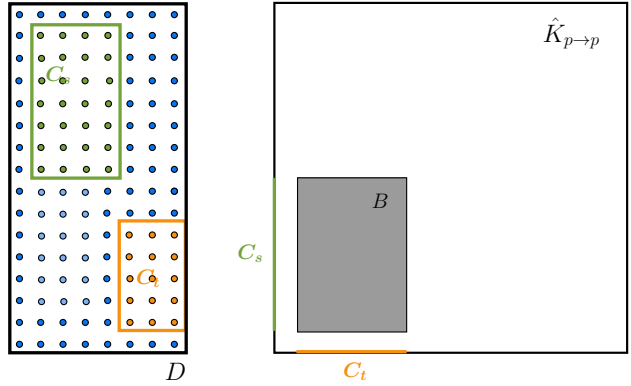}
    \centering
    \caption{On the left, two subdomains $C_s, C_t \subset D$ containing discrete points of the poloidal plane discretization. On the right, $B$ is the submatrix  of  $\hat{K}_{p \rightarrow p}$ that contains the evaluation of the integral operator along all paths that start from points in $C_t$ and end in points in $C_s$.}
    \label{fig:cl}
\end{figure}

\textbf{Low-rank Interactions.} 
 Let us consider two subdomains $C_s, C_t \subset D$ of the poloidal plane $D$, as indicated in Figure~\ref{fig:cl}. Let $B$ denote the submatrix that describes the interaction of the points contained in these two subdomains, that is, its row and column indices correspond to the grid points contained in $C_s$ and $C_t$, respectively.
 The entries of $B$ contain the evaluation of the kernel operator along paths that start from grid points in $C_t$ and end in grid points in 
 $C_s$; see also~\eqref{eq:discKpp}.
If $C_s$ and $C_t$ are very close or even overlap, some entries $B_{ij}$ correspond to the evaluation of the kernel in a region close to the singularity and, hence, $B$ has a slow singular value decay. On the other hand, if $C_s$ and  $C_t$ are well separated, the kernel is evaluated in regions where it is smooth, and the submatrix $B$ has a fast singular value decay, suggesting that these blocks can be compressed effectively.

\textbf{Admissibility Condition.} The above considerations motivate the definition of admissible and inadmissible blocks: The physical distance $\text{dist}( C_{s}, C_{t})$ between two subdomains $C_s,C_t$ determines whether the corresponding submatrix $B$ is amenable to low-rank approximation. This classification is based on so called admissibility criteria~\cite{Hackbusch2015}. We say that two subdomains $C_{s}$ and $C_{t}$ on the same level of the cluster tree are admissible if they satisfy
\begin{equation}
    \label{eq:admiss}
    \text{dist}( C_{s}, C_{t}) > 0
\end{equation}
and they are called inadmissible otherwise.
On level 1 of the cluster tree (see Figure~\ref{fig:2_1}), it turns out that all four submatrices are inadmissible because they have a common edge or vertex.
On level 2, a significant portion of the blocks becomes admissible (see Figure~\ref{fig:2_2}); these blocks are approximated by low-rank matrices. The remaining, inadmissible blocks
are partitioned again and the admissibility criterion is evaluated for the corresponding subsmatrices. This procedure is repeated for every inadmissible block until the last level of the cluster tree. In this way, the interaction of neighboring regions is presented on finer levels, while distant interactions are captured at coarser levels. In practice, 
the criterion of Eq.~\eqref{eq:admiss} is  evaluated by imposing a minimum distance of $\sqrt{2}$ on the grid points contained in the subdomains $C_s,C_t$.

\begin{figure}[H]
    \centering
    \begin{subfigure}[c]{0.3\textwidth}
        \centering
        \scalebox{0.57}{

\begin{tikzpicture}[x=0.75pt,y=0.75pt,yscale=-1,xscale=1]
%uncomment if require: \path (0,488); %set diagram left start at 0, and has height of 488

%Shape: Rectangle [id:dp19565439629087988] 
\draw  [draw opacity=0][fill={rgb, 255:red, 241; green, 185; blue, 103 }  ,fill opacity=0.67 ] (219,40) -- (174,40) -- (174,130) -- (219,130) -- cycle ;
%Shape: Rectangle [id:dp8587483916308851] 
\draw  [draw opacity=0][fill={rgb, 255:red, 131; green, 71; blue, 183 }  ,fill opacity=0.58 ] (174,40) -- (129,40) -- (129,130) -- (174,130) -- cycle ;
%Shape: Rectangle [id:dp3193425644571768] 
\draw  [draw opacity=0][fill={rgb, 255:red, 225; green, 75; blue, 78 }  ,fill opacity=0.88 ] (264,130) -- (219,130) -- (219,220) -- (264,220) -- cycle ;
%Shape: Rectangle [id:dp3588718333850651] 
\draw  [draw opacity=0][fill={rgb, 255:red, 32; green, 128; blue, 241 }  ,fill opacity=0.77 ] (219,310) -- (174,310) -- (174,400) -- (219,400) -- cycle ;
%Straight Lines [id:da25775091152667384] 
\draw [color={rgb, 255:red, 148; green, 208; blue, 78 }  ,draw opacity=1 ][fill={rgb, 255:red, 255; green, 255; blue, 255 }  ,fill opacity=1 ][line width=2.25]    (84,130) -- (264,130) ;
%Straight Lines [id:da2210179638697911] 
\draw [color={rgb, 255:red, 123; green, 179; blue, 245 }  ,draw opacity=1 ][fill={rgb, 255:red, 255; green, 255; blue, 255 }  ,fill opacity=1 ][line width=2.25]    (84,220) -- (264,220) ;
%Straight Lines [id:da834119381688457] 
\draw [color={rgb, 255:red, 148; green, 208; blue, 78 }  ,draw opacity=1 ][fill={rgb, 255:red, 255; green, 255; blue, 255 }  ,fill opacity=1 ][line width=2.25]    (84,310) -- (264,310) ;
%Straight Lines [id:da8015702060816815] 
\draw [color={rgb, 255:red, 148; green, 208; blue, 78 }  ,draw opacity=1 ][fill={rgb, 255:red, 255; green, 255; blue, 255 }  ,fill opacity=1 ][line width=2.25]    (129,40) -- (129,400) ;
%Straight Lines [id:da9737565104455929] 
\draw [color={rgb, 255:red, 123; green, 179; blue, 245 }  ,draw opacity=1 ][fill={rgb, 255:red, 255; green, 255; blue, 255 }  ,fill opacity=1 ][line width=2.25]    (174,40) -- (174,400) ;
%Straight Lines [id:da014459215494917466] 
\draw [color={rgb, 255:red, 148; green, 208; blue, 78 }  ,draw opacity=1 ][fill={rgb, 255:red, 255; green, 255; blue, 255 }  ,fill opacity=1 ][line width=2.25]    (219,40) -- (219,400) ;
%Shape: Rectangle [id:dp8000420561434548] 
\draw  [line width=2.25]  (84,40) -- (264,40) -- (264,400) -- (84,400) -- cycle ;

% Text Node
\draw (94.72,337.01) node [anchor=north west][inner sep=0.75pt]  [font=\huge,color={rgb, 255:red, 148; green, 208; blue, 78 }  ,opacity=1 ] [align=left] {1};
% Text Node
\draw (139.93,337.01) node [anchor=north west][inner sep=0.75pt]  [font=\huge,color={rgb, 255:red, 148; green, 208; blue, 78 }  ,opacity=1 ] [align=left] {2};
% Text Node
\draw (139.28,247) node [anchor=north west][inner sep=0.75pt]  [font=\huge,color={rgb, 255:red, 148; green, 208; blue, 78 }  ,opacity=1 ] [align=left] {3};
% Text Node
\draw (91,247) node [anchor=north west][inner sep=0.75pt]  [font=\huge,color={rgb, 255:red, 148; green, 208; blue, 78 }  ,opacity=1 ] [align=left] {4};
% Text Node
\draw (234.18,246.97) node [anchor=north west][inner sep=0.75pt]  [font=\huge,color={rgb, 255:red, 148; green, 208; blue, 78 }  ,opacity=1 ] [align=left] {7};
% Text Node
\draw (234.18,337.01) node [anchor=north west][inner sep=0.75pt]  [font=\huge,color={rgb, 255:red, 148; green, 208; blue, 78 }  ,opacity=1 ] [align=left] {6};
% Text Node
\draw (186.88,337.01) node [anchor=north west][inner sep=0.75pt]  [font=\huge,color={rgb, 255:red, 148; green, 208; blue, 78 }  ,opacity=1 ] [align=left] {5};
% Text Node
\draw (178.95,67) node [anchor=north west][inner sep=0.75pt]  [font=\huge,color={rgb, 255:red, 148; green, 208; blue, 78 }  ,opacity=1 ] [align=left] {12};
% Text Node
\draw (226.35,67.99) node [anchor=north west][inner sep=0.75pt]  [font=\huge,color={rgb, 255:red, 148; green, 208; blue, 78 }  ,opacity=1 ] [align=left] {11};
% Text Node
\draw (225.03,160.55) node [anchor=north west][inner sep=0.75pt]  [font=\huge,color={rgb, 255:red, 148; green, 208; blue, 78 }  ,opacity=1 ] [align=left] {10};
% Text Node
\draw (187,161) node [anchor=north west][inner sep=0.75pt]  [font=\huge,color={rgb, 255:red, 148; green, 208; blue, 78 }  ,opacity=1 ] [align=left] {9};
% Text Node
\draw (186.74,247) node [anchor=north west][inner sep=0.75pt]  [font=\huge,color={rgb, 255:red, 148; green, 208; blue, 78 }  ,opacity=1 ] [align=left] {8};
% Text Node
\draw (87.14,161) node [anchor=north west][inner sep=0.75pt]  [font=\huge,color={rgb, 255:red, 148; green, 208; blue, 78 }  ,opacity=1 ] [align=left] {13};
% Text Node
\draw (87,67) node [anchor=north west][inner sep=0.75pt]  [font=\huge,color={rgb, 255:red, 148; green, 208; blue, 78 }  ,opacity=1 ] [align=left] {16};
% Text Node
\draw (133.57,67) node [anchor=north west][inner sep=0.75pt]  [font=\huge,color={rgb, 255:red, 148; green, 208; blue, 78 }  ,opacity=1 ] [align=left] {15};
% Text Node
\draw (134.44,161) node [anchor=north west][inner sep=0.75pt]  [font=\huge,color={rgb, 255:red, 148; green, 208; blue, 78 }  ,opacity=1 ] [align=left] {14};

\end{tikzpicture}

}
        \vspace{6mm}
        \subcaption{\centering Subdivision of $D$ at level 2 of the cluster tree}
        \label{fig:3_1}
    \end{subfigure}
    \hspace{15mm}
    \begin{subfigure}[c]{0.51\textwidth}
        \centering
        \includegraphics[width=\textwidth]{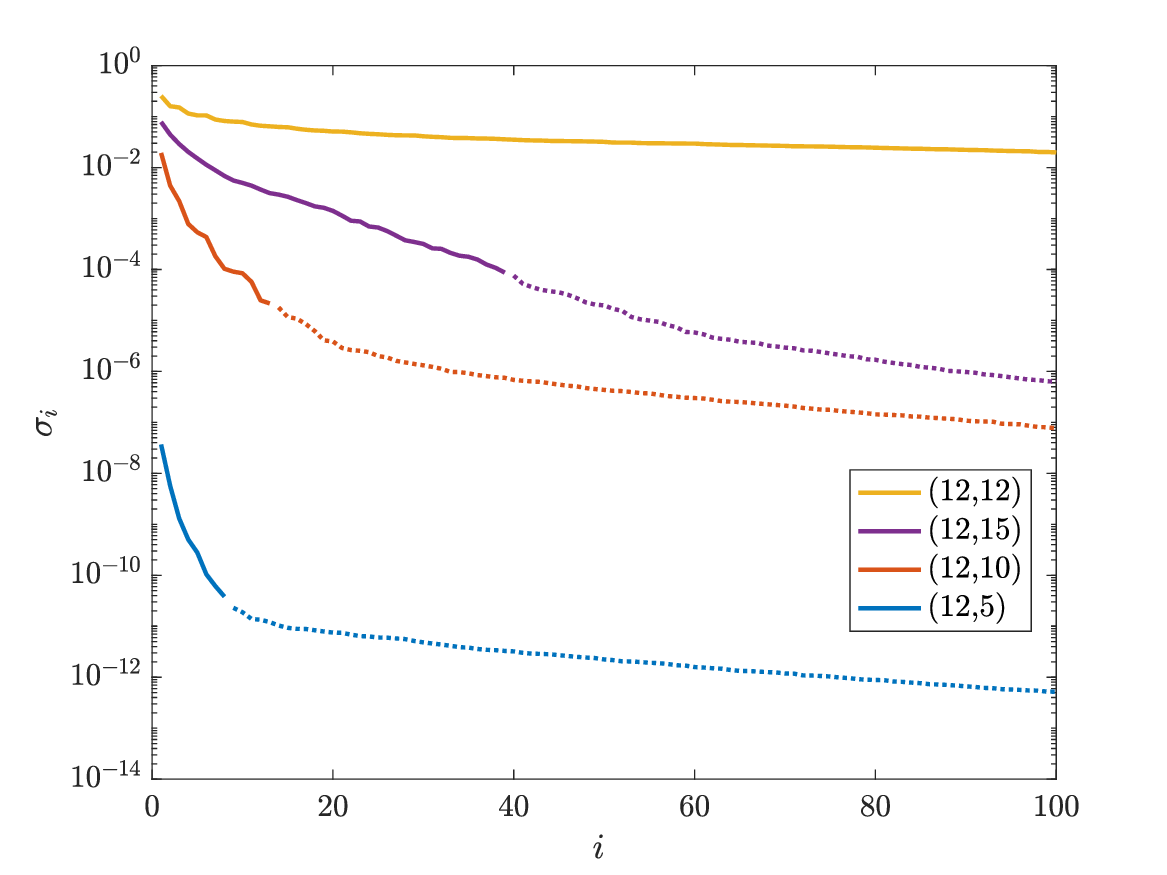}
        \centering
        \subcaption{\centering First 100 singular values of $300 \times 300$ submatrices}
        \label{fig:3_2}
    \end{subfigure} 
    \centering
    \caption{Singular value decay for the submatrices corresponding to the interaction of subdomain 12 with subdomains 12, 15, 10 and 5.}
    \label{fig:3}
\end{figure}
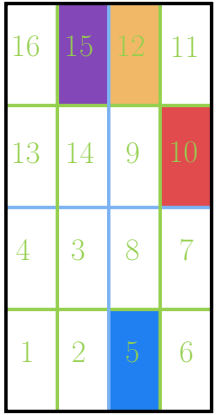
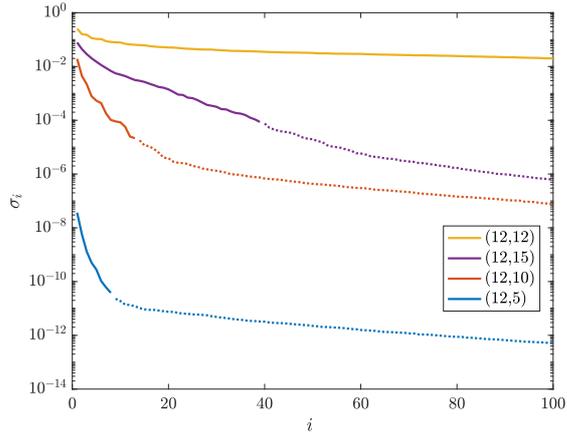

\textbf{Approximability of $\hat{K}_{p \rightarrow p}$.} To demonstrate the efficacy of low-rank 
approximation for admissible blocks, we examine the singular value decay for some of the submatrices of $\hat{K}_{p \rightarrow p}$ at level 2. Specifically, we analyze the interactions of subdomain 12 (highlighted in Figure~\ref{fig:3_1}) with: itself, subdomain 15 (which shares an edge), subdomain 10 (which shares a vertex), and subdomain 5 (a well-separated subdomain). Note that admissibility criterion~\eqref{eq:admiss} excludes interactions between subdomains that share an edge or a vertex. 
Figure~\ref{fig:3_2}  displays the first 100 singular values of the corresponding $300 \times 300$ submatrices. The results show that the block associated with the two well-separated subdomains exhibits rapid singular value decay and thus admits an excellent low-rank representations. When the subdomains share a vertex, the singular value decay deteriorates somewhat and this deterioration becomes more pronounced when they share an edge. Finally, the block corresponding to self-interaction
does not feature any practically useful singular value decay. In Figure~\ref{fig:3_2}, singular values below $10^{-3} \times \sigma_1$ are shown as dotted lines. Note that
the singular value decay slows down significantly below that level, due to the numerical error introduced by the interpolation of the plasma model solution during the construction of the matrix $\hat{K}_{p \rightarrow p}$.
%However, this condition could be relaxed to allow vertex-sharing subdomains to be considered admissible, potentially increasing the number of blocks that benefit from low-rank approximations.

\textbf{Adaptive Cross Approximation.} While Figure~\ref{fig:3_2} confirms the presence of low-rank structure in admissible blocks, computing the corresponding approximation would require the singular value decomposition (SVD) of the \emph{fully} assembled block.
In order to avoid this and perform low-rank approximations by assembling only a fraction of the matrix entries,
we use Adaptive Cross Approximation (ACA)~\cite{Bebendorf2000,Bebendorf2003}. This approach iteratively constructs a rank-$r$ approximation from $r$ selected rows and columns.

In the first step, ACA chooses a pivot pair $(i_1,j_1)$ and computes a rank-1 approximation
\[
 A \approx \boldsymbol{u_1} \boldsymbol{v_1}^T, \quad \boldsymbol{u_1} = A(:,j_1), \quad  \boldsymbol{v_1} = A(i_1,:) / A(i_1,j_1),
\]
where use Matlab's colon notation: $A(:,j_1)$ denotes column $j_1$ and $A(i_1,:)$ denotes row $i_1$ of $A$.
This process is repeated for the residual matrix $R_1 = A - \boldsymbol{u_1} \boldsymbol{v_1}^T$ in order to obtain the second term 
in the rank-two approximation
$A \approx \boldsymbol{u_1} \boldsymbol{v_1}^T + \boldsymbol{u_2} \boldsymbol{v_2}^T$. Then the corresponding residual $R_2 = R_1 - \boldsymbol{u_2} \boldsymbol{v_2}^T$
is considered, and so on. Each step relies on the selection of an index pair $(i_k,j_k)$ such that the pivot element $R_{k-1}(i_k,j_k)$ with $R_0 = A$ is not too small. Ideally, this pivot element should be chosen as large as possible in magnitude, but this is impractical because it would require evaluating the entire matrix $A$. Instead,
the first index $i_1$ is chosen arbitrarily, while all subsequent indices are determined by locally maximizing within the currently selected row/column.

\begin{algorithm}[H]
\footnotesize
\caption{ACA with partial pivoting}
\begin{algorithmic}[1]
\Require Matrix \( A \), tolerance \( \varepsilon>0 \)
\Ensure Vectors $\boldsymbol{u_k}, \boldsymbol{v_k}$ defining rank-$r$ approximation $A \approx \boldsymbol{u_1} \boldsymbol{u_1}^T +\cdots + \boldsymbol{u_r} \boldsymbol{u_r}^T$
\State Initialize: \( R_0 = A \), \( k = 0 \), \( I = \{\} \), \( J = \{\} \), \( i_k = 1 \)
\Repeat
    \State \( j_k = \arg\max_j |R_{k-1}(i_k,j)| \)
    \State \( \boldsymbol{u_k} = R_{k-1}(:,j_k) \)
    \State \( \boldsymbol{v_k} = R_{k-1}(i_k,:)^{T} / R_{k-1}(i_k,j_k) \)
    \State \( R_k = R_{k-1} - \boldsymbol{u_k} \boldsymbol{v_k}^T \)
    \State \( k = k+1 \)
    \State \( I \leftarrow I \cup \{i_k\}, \quad J \leftarrow J \cup \{j_k\} \)
    \State \( i_k = \arg\max_{i \notin I} |u_k(i)| \)
\Until{Stopping criterion is reached for tolerance $\varepsilon$}
\end{algorithmic}
\label{alg:ACA}
\end{algorithm}
The resulting method is summarized in Algorithm~\ref{alg:ACA}. Let us stress that the residuals $R_k = A - \boldsymbol{u_1} \boldsymbol{v_1}^T \cdots - \boldsymbol{u_k} \boldsymbol{v_k}^T$ are never actually formed; the entries of $R_k$ required in Algorithm~\ref{alg:ACA} are computed on-the-fly by evaluating and updating the corresponding entries of $A$. Thus, only $r(m+n)$ entries of $A$ need to be evaluated in the course of the algorithm.

An important aspect of ACA is choosing a suitable stopping criterion that does not require the evaluation of the whole residual. Following the approximation proposed in~\cite{Bebendorf2003}, where the norm of the residual after $k$ steps is estimated as $\| R_k \|_2 \approx \|\boldsymbol{u_k}\|_2 \|\boldsymbol{v_k}\|_2$, Algorithm~\ref{alg:ACA} is stopped when
\begin{equation}
    \|\boldsymbol{u_k}\|_2 \|\boldsymbol{v_k}\|_2\leq \varepsilon \|\boldsymbol{u_0}\|_2 \|\boldsymbol{v_0}\|_2
    \label{eq:stopcr}
\end{equation}
is satisfied for the prescribed tolerance $\varepsilon$.
In other words, when the norm of the next rank-1 term falls below a prescribed tolerance, the process is terminated. This ensures that the approximation remains accurate while avoiding unnecessary computations. %If at any point the pivot element is zero, meaning no significant updates can be made, the algorithm halts early. 

% When the decay of the error is slow, additional steps may be required, or alternative pivoting strategies can be considered to enhance convergence.

The computational complexity of ACA algorithm is \(\mathcal{O}(r^2(m+n))\), which is significantly lower than computing the SVD.
Both, the memory complexity and the computational complexity for a matrix-vector multiplication using the ACA approximation are \(\mathcal{O}(r(m+n))\), making it an attractive alternative to dense matrix operations.

In the context of our work, ACA is used to compute a low-rank approximation for each admissible block in the hierarchical matrix representation of $\hat{K}_{p \rightarrow p}$. We begin by constructing a cluster tree through recursive partitioning of the physical domain. Using the admissibility criterion~\eqref{eq:admiss}, we classify matrix blocks into admissible and inadmissible. Blocks that remain inadmissible on the lowest tree level correspond to near-field interactions and capture the kernel singularity; they are stored as dense matrices. For admissible blocks, ACA is applied to compute a low-rank approximation, which is stored via the two low-rank factors. The resulting hierarchical matrix is stored as a tree data structure, which enables fast matrix-vector operations without assembling the full dense matrix. Assuming that the ranks are constant, the low-rank factors of all admissible blocks on each level of the recursion require the evaluation and storage of $\mathcal O(N_P)$ entries. After $\mathcal O(\log N_P)$ recursions, there are $\mathcal O(N_P)$ inadmissible blocks of constant block size, which are stored as dense matrices. In summary, storing and multiplying with the hierarchical low-rank approximation of the $N_P\times N_P$ matrix $\hat{K}_{p \rightarrow p}$ has complexity $\mathcal O(N_P \log N_P)$.

\section{Numerical Results}
\label{sec:numres}

The simulations considered in this work are carried out with
the GBS code~\cite{Giacomin2022a},~\cite{Ricci2012b}, a three-dimensional, flux-driven code used to
study plasma turbulence in the tokamak boundary, where the self-consistent kinetic neutral model described in Section~\ref{sec:neutralmodel}, is coupled with the drift-reduced Braginskii equations~\cite{Braginskii1965} to model the plasma.

\textbf{Simulations Setup.} The simulation setup is given by TCV-X21 tokamak discharge~\cite{Oliveira2022}, an experimental dataset developed for a validation studies of turbulence codes on the TCV tokamak. TCV-X21 is a lower single-null L mode discharge performed at low toroidal magnetic field. The domain considered for our tests corresponds approximately to half the size of the TCV tokamak. In practice, the computational grid for the solution of the plasma model is $N_{x} \times N_{y} \times N_{\parallel} = 150 \times 300 \times 64$. The neutral model is solved every 2000 plasma time steps, typically on a coarser spatial grid, and the resulting neutral moments are interpolated to the plasma grid using two-dimensional piecewise linear interpolation. All results in this section are shown on the plasma grid ($N_x \times N_y = 150 \times 300$) for a fixed poloidal plane.
For the purpose of our tests, we disable reflected trajectories by setting $\alpha_{\textrm{refl}} = 0$ and neglect both recombination and gas puffing. Although these simplifications reduce the physical fidelity of the model, they do not influence the assessment of the numerical performance of the hierarchical matrix method. Additional physical mechanisms are supported by our implementation and can be reactivated as needed.

\textbf{Hardware and Parallelization Strategy.} GBS is implemented in Fortran 90, relying on the PETSc library~\cite{Balay1998} for the linear solvers and  Intel MPI 19.1 for the parallelization. 
Simulations were performed on the Eiger supercomputer at the Swiss National Supercomputing Centre (CSCS). Eiger is a Cray EX system equipped with AMD EPYC 7742 processors. Each compute node features two 64-core AMD EPYC 7742 CPUs (2.25 GHz), for a total of 128 physical cores and 256 GB of memory per node. 
To evolve the plasma model of the considered simulations, 128 MPI tasks per node are employed across 8 nodes, totaling 1024 parallel tasks. The domain decomposition strategy leverages the geometry of the problem: the toroidal direction is parallelized using one MPI task per toroidal plane, while the grid on each poloidal plane is further split across 16 MPI tasks  using domain decomposition. This strategy enables efficient distribution of both the computational load and memory footprint across the allocated resources.

 The goal of the numerical tests is to demonstrate that the hierarchical matrix method improves simulation efficiency while maintaining accuracy and enabling resolutions that were previously computationally infeasible.

\textbf{Dense Solver.}
As  baseline method for the neutral model solution, we consider the one described in~\cite{Giacomin2022}, where we assembly the whole dense matrix in Eq.~\eqref{eq:linsys} and solve the linear system using  unpreconditioned GMRES from PETSc, with a tolerance on the residual of $10^{-10}$. 
The matrix is distributed in parallel using the same MPI layout as the plasma solver: each poloidal plane is parallelized via domain decomposition.

\textbf{Hierarchical Matrix Solver.}
The hierarchical matrix approximation of $\hat{K}_{p \rightarrow p}$ described in Section~\ref{sec:hm} is implemented in Fortran as part of the GBS code. The implementation of the data structure assembly, including the low-rank approximation via ACA, have been implemented from scratch. The tolerance used for the stopping criterion of ACA~\eqref{eq:stopcr} is $\varepsilon = 10^{-3}$. The implementation of the matrix-vector operation for hierarchical matrices exploit BLAS routines for local dense matrix-vector products. This is achieved by relying on the \texttt{MatShell} feature of PETSc, thus avoiding explicit matrix assembly when carrying unpreconditioned GMRES for solving~\eqref{eq:linsys}. The GMRES solver uses the same residual tolerance of $10^{-10}$ as in the dense case. For the hierarchical matrix solver, only the toroidal direction is parallelized, assigning one MPI task per poloidal plane.

 In the remaining part of this section, we refer to results obtained using the hierarchical matrix solver as \textit{HM}, and those using the dense solver as \textit{Dense}.

\begin{figure}[ht]
    \centering
    \begin{subfigure}[t]{0.32\textwidth}
        \centering
        \includegraphics[width=0.85\textwidth]{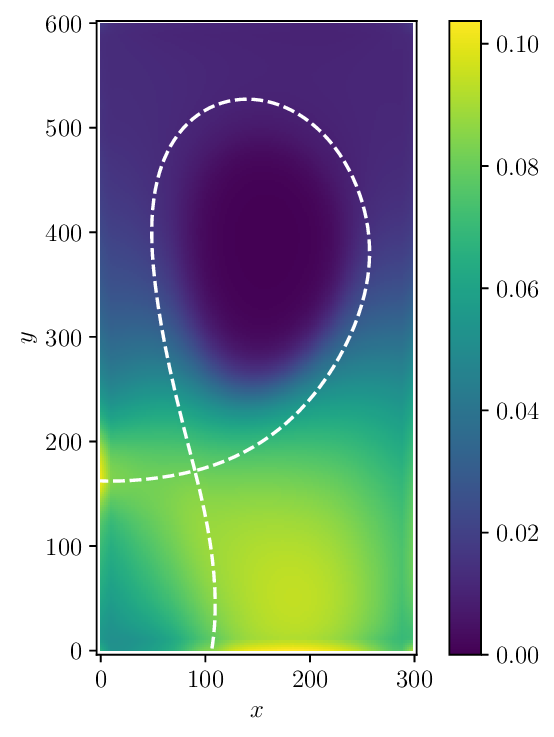}
        \centering
        \subcaption{Dense, $N_{x} \times N_{y} = 40 \times 80 $}
        %\label{fig:5_1}
    \end{subfigure}
    \hfill
    \begin{subfigure}[t]{0.32\textwidth}
        \centering
        \includegraphics[width=0.85\textwidth]{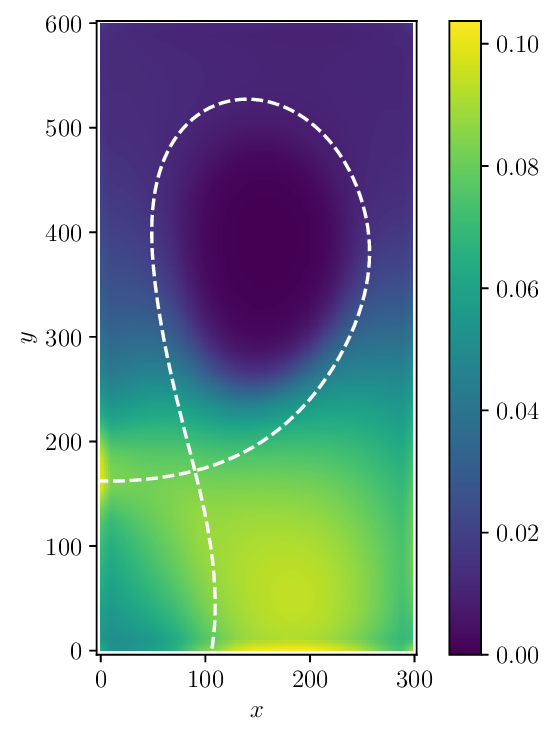}
        \centering
        \subcaption{HM, $N_{x} \times N_{y} = 40\times80$}
        % \label{fig:2_1}
    \end{subfigure}
    \hfill
    \begin{subfigure}[t]{0.32\textwidth}
        \centering
        \includegraphics[width=0.90\textwidth]{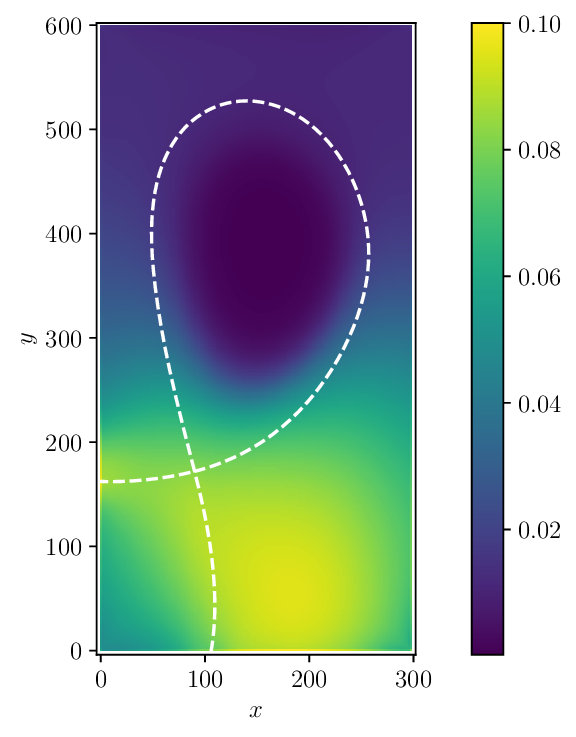}
        \centering
        \subcaption{ HM, $N_{x} \times N_{y} = 150\times300$}
        % \label{fig:2_2}
    \end{subfigure} 
    \centering
    \caption{Neutral density profile on one poloidal plane, interpolated on the plasma grid $N_{x} \times N_{y} = 150\times300$}
    \label{fig:5}
\end{figure}

\textbf{High-resolution solution.} We begin by validating the HM method against the dense solver using neutral density profiles. Figure~\ref{fig:5} shows the neutral density on a single poloidal plane for (a) the dense solver on a $40 \times 80$ grid, (b) the HM solver on the same grid, and (c) the HM solver on a higher-resolution $150 \times 300$ grid, corresponding to the plasma grid. 

We highlight that the memory required by GBS for the interpolation and setup phase, in addition to the linear system assembly and solution, makes the treatment with the dense method infeasible for the high resolution $150 \times 300$, exceeding the total memory available per node.
Therefore, the result in Figure~\ref{fig:5}(c) represents the first time the neutral density profile is computed and visualized at the full plasma resolution. The enhanced resolution reveals finer boundary layer structures near the targets, which are not captured at coarser grid levels.

\begin{figure}[H]
    \centering
    \begin{subfigure}[t]{0.49\textwidth}
        \centering
        \includegraphics[width=0.82\textwidth]{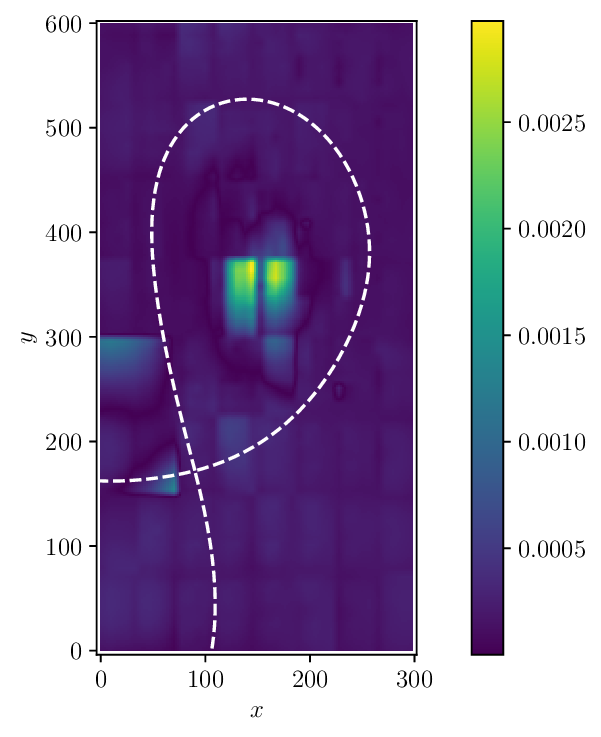}
        \subcaption{Relative error of HM neutral density against the dense solution, both computed on a grid $N_{x} \times N_{y} = 40 \times 80$.}
        \label{fig:6_1}
    \end{subfigure}
    \hfill
    \begin{subfigure}[t]{0.49\textwidth}
        \centering
        \includegraphics[width=0.80\textwidth]{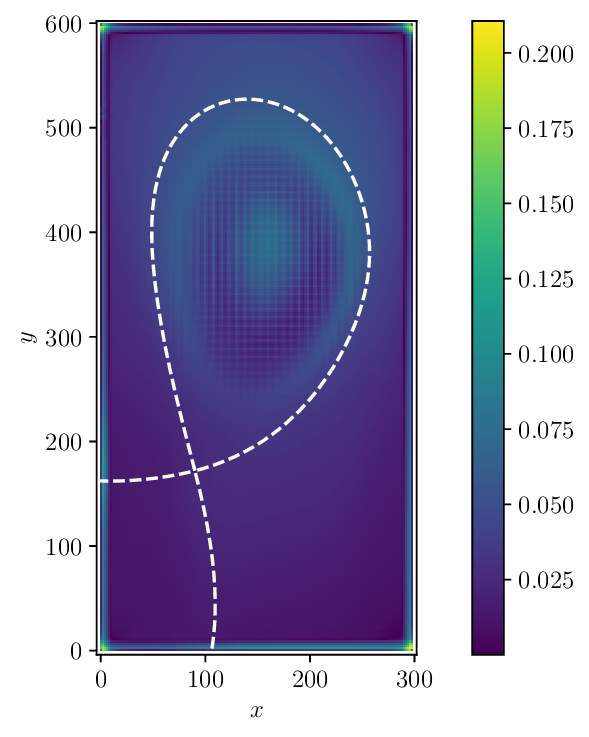}
        \subcaption{Relative error of HM neutral density computed on the high-resolution grid $N_{x} \times N_{y} = 150\times 300$, against the dense solution, computed on a grid $N_{x} \times N_{y} = 40 \times 80$. }
        \label{fig:6_2}
    \end{subfigure} 
    \caption{Relative errors computed using the average density profile over the 64 poloidal planes, and over 4 GBS time units.}
    \label{fig:6}
\end{figure}

Figure~\ref{fig:6} shows relative errors between the dense and hierarchical matrix solutions, computed using the average density profile over the 64 poloidal planes, and over 4 GBS time units.

In particular, Figure~\ref{fig:6}(a) shows that the relative error of the HM solution with respect to the dense one, for the $40\times80$ grid, remains below 0.25\% across the entire domain. This confirms the accuracy of the low-rank approximations we have performed imposing a ACA low-rank approximation tolerance of $\varepsilon = 10^{-3}$.
In Figure~\ref{fig:6}(b) we report the relative error of the hierarchical $150\times300$ solution with respect to the $40\times80$ one, showing that the finer grid of  $150\times300$ enabled by the hierarchical matrix method  improves the resolution of the boundary layers  at the targets.  Indeed, the relative error attains a maximum of 20\% in the left and bottom wall boundary region.
It is worth noticing that, in the computation of the hierarchical matrix approximation for the coarse grid, 27\% of the $\hat{K}_{p \rightarrow p}$ matrix entries are computed, while for the finer grid,  only 3\% of the elements of the matrix are evaluated. 

\begin{figure}[H]
    \centering
    \includegraphics[width=0.6\textwidth]{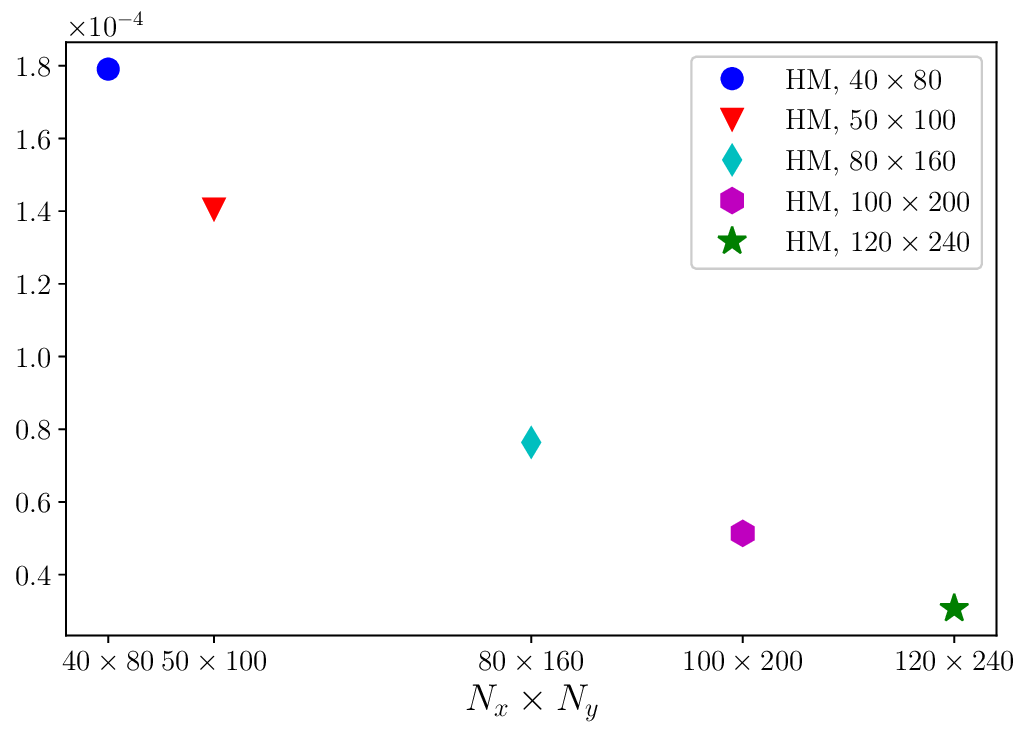} 
    \centering
    \caption{Convergence of the L2 norm of the HM solution toward the plasma model resolution ($150 \times 300$)}
    \label{fig:7}
\end{figure}
    
To confirm that the discrepancies in Figure~\ref{fig:6_2} stem from improved resolution rather than modeling error, Figure~\ref{fig:7} shows convergence of the $L_2-$norm of the HM solution as the neutral grid is refined, towards the hierarchical matrix approximation solution computed on the $150 \times 300$ grid.

\begin{figure}[ht]
    \centering
    \begin{subfigure}[t]{0.48\textwidth}
        \centering
        \includegraphics[width=0.9\textwidth]{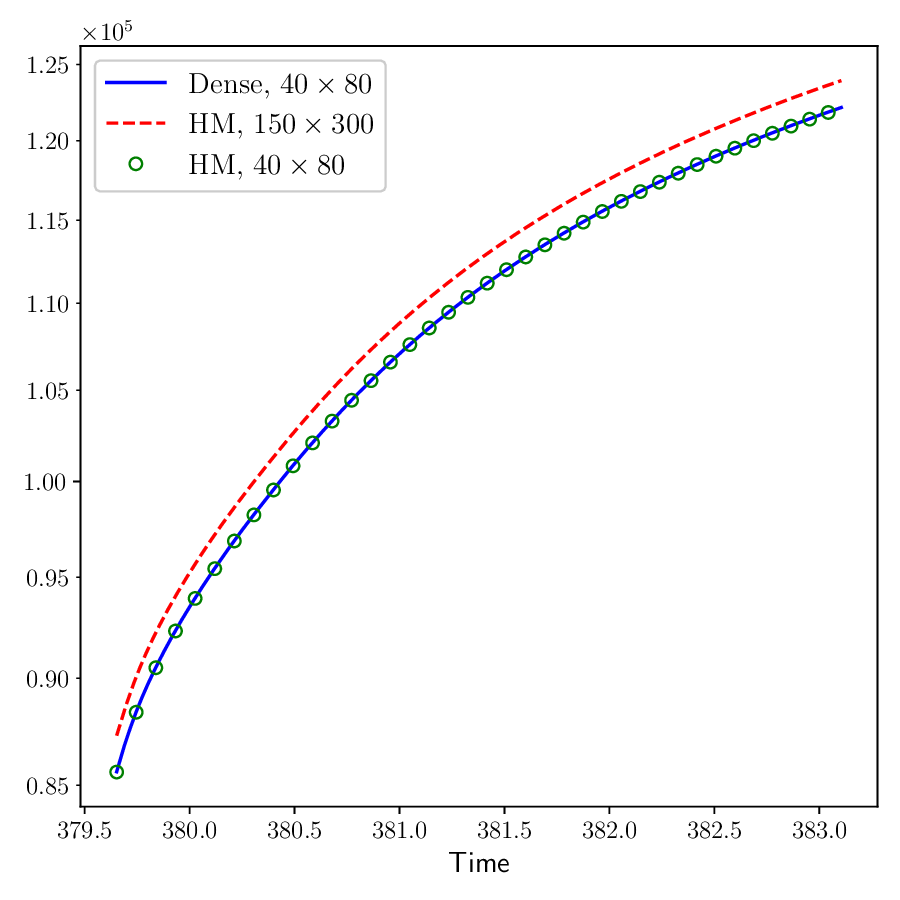}
        \subcaption{ }
        \label{fig:8_1}
    \end{subfigure}
    \hfill
    \begin{subfigure}[t]{0.48\textwidth}
        \centering
        \includegraphics[width=0.9\textwidth]{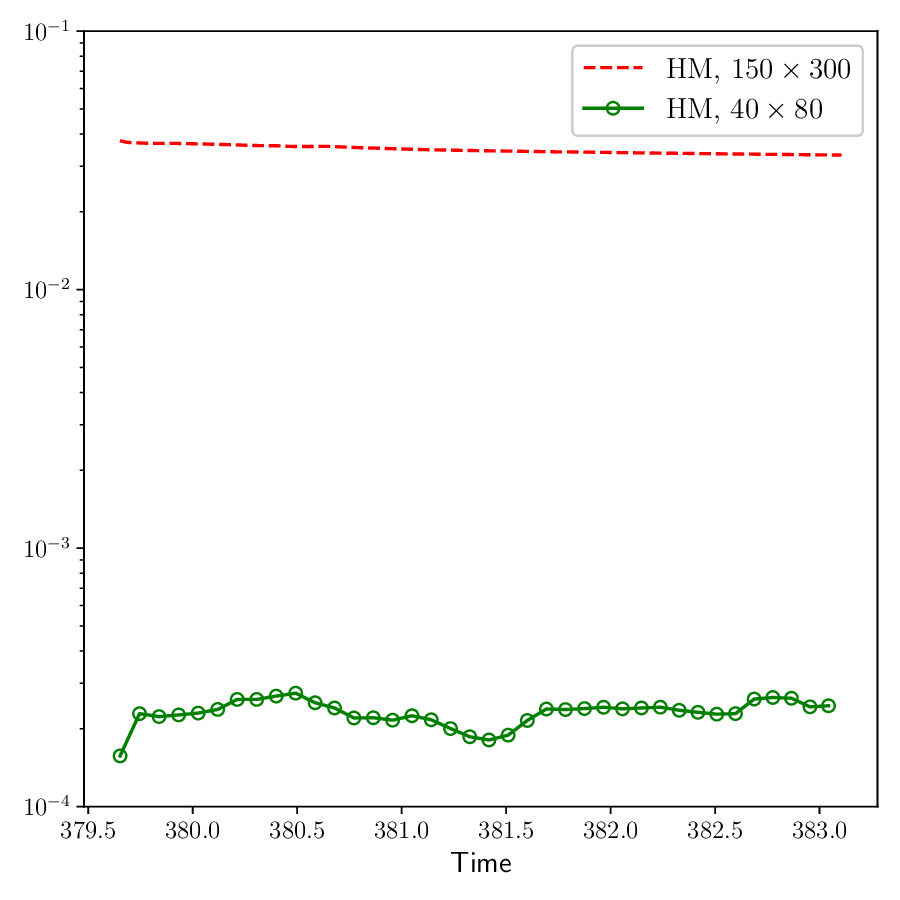}
        \centering
        \subcaption{ }
        \label{fig:8_2}
    \end{subfigure} 
    \centering
    \caption{ Comparison of neutral density evolution between coarse-grid dense and fine-resolution HM simulations. In (a) total neutral density sum over the domain and in (b) the average of the relative error with respect to the dense case, computed locally for each point of the domain. The $y$ axis  is in logarithmic scale.}
    \label{fig:8}
\end{figure}

In order to assess the implications of the finer resolutions on the results of the simulations, we evolve the plasma and neutral models for 4 GBS simulations time units, corresponding to approximately $40 \mu s$. To analyze the results, we compute the sum of the neutral density values over the entire three-dimensional domain for the coarse and fine resolution hierarchical matrix solutions, as well as for the coarse grid dense solution, as shown in Figure~\ref{fig:8_1}. We observe that for the coarse grid the results is the same as the dense case, as expected, while for the finer resolution, the total density sum is higher.

In Figure~\ref{fig:8_2}, for every timestep (sampled every 2500 steps), we compute the relative error at each point in the domain with respect to the coarse dense solution and then averaged over the domain. When comparing the hierarchical matrix and dense solutions on the same coarse grid ($40 \times 80$), the relative difference stabilizes around $10^{-4}$, which is below  the tolerance of the hierarchical method and confirms that the discrepancy is purely due to the approximation error. On the other hand, when increasing the resolution in the hierarchical matrix solution ($150 \times 300$ versus dense $40 \times 80$), the relative difference is larger, and stable around $10^{-2}$ as the finer grid is able to better resolve the target regions, thus capturing more details in the neutral density distribution.

\begin{table}[H]
    \setlength{\tabcolsep}{7pt} % Default value: 6pt
    \renewcommand{\arraystretch}{1.3} % Default value: 1
    
    \centering
    \begin{tabular}{|l|l|ll|ll|l|}
        \multicolumn{1}{l}{}    & \multicolumn{1}{l}{} & \multicolumn{2}{l|}{Time for $\hat{K}_{p \rightarrow p}$ [s]} & \multicolumn{2}{l}{GMRES time [s]} & \multicolumn{1}{l}{}  \\ 
        \hline
        \textbf{$N_x \times N_y$} & \textbf{$N_P$}          & Dense & HM                                                                   & Dense & HM                                & \textbf{Memory load}  \\ 
        \hline
        $50 \times 100$         & 5000                 & 24.2   & 6.8                                                                   & 0.5   & 0.2                              & $27.2 \,\%$           \\ 
        \hline
        $80 \times 160$         & 12800                & 158.5  & 20.5                                                                   & 8.2  & 1.3                             & $12.4 \, \%$           \\ 
        \hline
        $100 \times 200$        & 20000                & 396  & 33.8                                                                  & 12.1    & 2                                 & $8.4 \,\%$            \\
        \hline
        $150 \times 300$        & 45000                & /  & 92                                                                  &  /   & 4.3                                & $4.2 \,\%$            \\
        \hline
    \end{tabular}
    \caption{Timings for assemblying the matrix $\hat{K}_{p \rightarrow p}$ and solving the linear system using unpreconditioned GMRES, for the dense and hierarchical matrix solver. The last column shows the percentage of matrix elements that are evaluated and stored in the hierarchical matrix case, with respect to the dense one. }
    \label{tab:timings}
    \end{table}

\textbf{Performance Improvement.} We now assess the performance benefits of the hierarchical matrix approach. Table~\ref{tab:timings} reports the time required to assemble $\hat{K}_{p \rightarrow p}$ and for the linear system solution with GMRES, for different neutral grid sizes, comparing  the dense and HM solver. The last column shows the memory load, defined as the percentage of matrix elements computed to construct the HM approximation of $\hat{K}_{p \rightarrow p}$. 
We compare the hierarchical matrix implementation with the dense version executed in serial.
We observe that the dense solver becomes infeasible at higher resolutions due to excessive memory requirements, while the HM solver enables simulations up to a grid size of $150 \times 300$. 

\begin{figure}[H]
    \centering
    \includegraphics[width=0.6\textwidth]{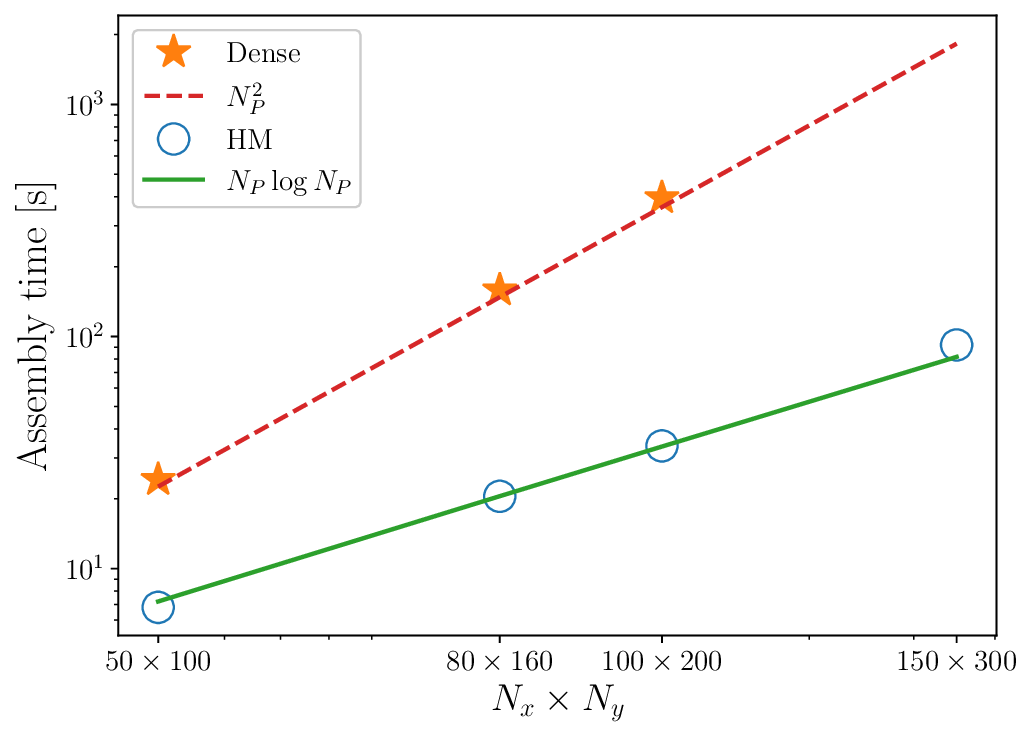} 
    \caption{Assembly time for the $N_P\times N_P$ matrix $\hat{K}_{p \rightarrow p}$ vs different grid sizes on logarithmic scale, with $N_P= N_x \cdot N_y$.}
    \label{fig:10}
\end{figure}      

Figure~\ref{fig:10} shows the asymptotic behavior of the assembly time as the neutral grid size increases. For a neutral grid of size $N_x \times N_y$, the matrix $\hat{K}_{p \rightarrow p}$ has dimension $N_P = N_x N_y$. The dense assembly time scales as $\mathcal{O}(N_P^2)$, as expected, whereas the hierarchical matrix assembly scales as $\mathcal{O}(N_P \log N_P)$, confirming our complexity estimate from Section~\ref{sec:hm}.

\section{Conclusions}

In this work, we accelerate the solution of kinetic neutral models used in the simulation of boundary plasma turbulence in tokamaks, when the kinetic equations for neutral particles are solved deterministically using the method of characteristics.
The bottleneck of the numerical solution is found in the assembly and solution of a dense linear system, resulting from the solution of an integral equation with a non-local and data-based kernel. Our approach employs a hierarchical matrix approximation to avoid the dense matrix assembly, enabling a significant reduction in the computational cost of forming and solving the resulting linear system, as well as the memory usage. 

Numerical tests based on the GBS code demonstrate that the hierarchical approach achieves speedups of one to two orders of magnitude in matrix assembly and solution times, without compromising the physical fidelity of the results. This enables simulations at resolutions that were previously infeasible with the dense solver, thereby improving the accuracy of the neutral model.

This study opens the door to efficient large-scale kinetic neutral simulations. Future work will explore parallel hierarchical matrix implementations and extend the approach to multispecies neutral models. With this method, the deterministic solution of the kinetic neutral model becomes a competitive alternative to the traditionally used Monte Carlo methods. Moreover, this approach will be essential for enabling high-fidelity kinetic neutral simulations in next-generation fusion devices of larger dimensions.

\vspace{3mm}

\textbf{Acknowledgments.} The authors thank Emmanuel Lanti and Brenno De Lucca for the helpful discussions regarding the GBS implementation of this work.

The simulations presented herein were carried out at CSCS (Swiss National Supercomputing Center).

This work has been carried out within the framework of the EUROfusion Consortium, via the Euratom Research and Training Programme (Grant Agreement No 101052200 — EUROfusion) and funded by the Swiss State Secretariat for Education, Research and Innovation (SERI). Views and opinions expressed are however those of the author(s) only and do not necessarily reflect those of the European Union, the European Commission, or SERI. Neither the European Union nor the European Commission nor SERI can be held responsible for them.

\textbf{Data Availability Statement.} The data that support the findings of this study are available upon reasonable request from the authors.

\printbibliography 

@BOOK{Golub2013,
  title = {Matrix computations},
  publisher = {Johns Hopkins University Press, Baltimore, MD},
  year = {2013},
  author = {Golub, G. H. and Van Loan, C. F.},
  edition = {Fourth},
}

@article{Baudoin2018,
author = {Baudoin, Camille and Tamain, Patrick and Bufferand, Hugo and Ciraolo, Guido and Fedorczak, Nicolas and Galassi, Davide and Ghendrih, Philippe and Nace, Nicolas},
doi = {10.1002/ctpp.201700168},
issn = {15213986},
journal = {Contrib. to Plasma Phys.},
number = {6-8},
pages = {484--489},
title = {{Turbulent heat transport in TOKAM3X edge plasma simulations}},
volume = {58},
year = {2018}
}

@article{Bebendorf2000,
author = {Bebendorf, M.},
issn = {0029-599X},
journal = {Numer. Math.},
pages = {565--589},
title = {{Approximation of boundary element matrices}},
volume = {86},
year = {2000}
}

@article {Bebendorf2003,
    AUTHOR = {Bebendorf, M. and Rjasanow, S.},
     TITLE = {Adaptive low-rank approximation of collocation matrices},
   JOURNAL = {Computing},
  FJOURNAL = {Computing. Archives for Scientific Computing},
    VOLUME = {70},
      YEAR = {2003},
    NUMBER = {1},
     PAGES = {1--24},
      ISSN = {0010-485X,1436-5057},
   MRCLASS = {65R20 (41A63 65F05 65F30)},
  MRNUMBER = {1972724},
       DOI = {10.1007/s00607-002-1469-6},
}

@article{Bufferand2017,
author = {Bufferand, H. and Baudoin, C. and Bucalossi, J. and Ciraolo, G. and Denis, J. and Fedorczak, N. and Galassi, D. and Ghendrih, Ph and Leybros, R. and Marandet, Y. and Mellet, N. and Morales, J. and Nace, N. and Serre, E. and Tamain, P. and Valentinuzzi, M.},
doi = {10.1016/j.nme.2016.11.031},
issn = {23521791},
journal = {Nucl. Mater. Energy},
pages = {852--857},
publisher = {Elsevier Ltd},
title = {{Implementation of drift velocities and currents in SOLEDGE2D–EIRENE}},
volume = {12},
year = {2017}
}

@phdthesis{Coroado2021,
author = {Calado Coroado, Andr{\'{e}}},
title = {Self-consistent simulation of multi-component plasma turbulence and neutral dynamics in the tokamak boundary},
school = {EPFL},
year = {2021}
}

@phdthesis{Giacomin2022,
author = {Giacomin, M.},
booktitle = {Ph.D. Thesis},
number = {Lausanne},
title = {Turbulent transport regimes in the tokamak boundary},
school = {EPFL},
year = {2022}
}

@article{Giacomin2022a,
author = {Giacomin, M. and Ricci, P. and Coroado, A. and Fourestey, G. and Galassi, D. and Lanti, E. and Mancini, D. and Richart, N. and Stenger, L. N. and Varini, N.},
doi = {10.1016/j.jcp.2022.111294},
journal = {J. Comput. Phys.},
title = {{The GBS code for the self-consistent simulation of plasma turbulence and kinetic neutral dynamics in the tokamak boundary}},
volume = {463},
year = {2022}
}

@article{Greengard1997,
author = {Greengard, L. and Rokhlin, V.},
doi = {10.1006/jcph.1997.5706},
issn = {00219991},
journal = {J. Comput. Phys.},
number = {2},
pages = {280--292},
title = {{A fast algorithm for particle simulations}},
volume = {135},
year = {1997}
}

@book{Hackbusch2015,
author = {Hackbusch, Wolfgang},
isbn = {9783662473238},
issn = {0179-3632},
publisher = {Springer Berlin, Heidelberg},
title = {{Hierarchical Matrices: Algorithms and Analysis}},
volume = {49},
year = {2015}
}

@book{Bebendorf2008,
author = {M. Bebendorf},
doi = {https://doi.org/10.1007/978-3-540-77147-0},
isbn = {9783540771463},
publisher = {Springer Berlin, Heidelberg},
title = {Hierarchical Matrices},
year = {2008}
}

@article{Reiter2005,
author = {Reiter, D. and Baelmans, M. and B{\"{o}}rner, P.},
doi = {10.13182/FST47-172},
issn = {15361055},
journal = {Fusion Sci. Technol.},
number = {2},
pages = {172--186},
title = {{The eirene and B2-eirene codes}},
volume = {47},
year = {2005}
}

@article{Ricci2012b,
author = {Ricci, P. and Halpern, F. D. and Jolliet, S. and Loizu, J. and Mosetto, A. and Fasoli, A. and Furno, I. and Theiler, C.},
doi = {10.1088/0741-3335/54/12/124047},
issn = {07413335},
journal = {Plasma Phys. Control. Fusion},
number = {12},
title = {{Simulation of plasma turbulence in scrape-off layer conditions: The GBS code, simulation results and code validation}},
volume = {54},
year = {2012}
}

@book{Sauter2010,
author = {Sauter, Stefan A. and Schwab, Christoph},
doi = {10.1007/978-3-540-78862-1},
publisher = {Springer Berlin},
title = {{Boundary Element Methods}},
url = {https://www.springer.com/gp/book/9783540306634},
year = {2010}
}

@article{Tamain2014,
author = {Tamain, P. and Bufferand, H. and Ciraolo, G. and Colin, C. and Ghendrih, Ph and Schwander, F. and Serre, E.},
doi = {10.1002/ctpp.201410017},
issn = {15213986},
journal = {Contrib. to Plasma Phys.},
number = {4-6},
pages = {555--559},
title = {{3D Properties of Edge Turbulent Transport in Full-Torus Simulations and their Impact on Poloidal Asymmetries}},
volume = {54},
year = {2014}
}

@phdthesis{Wersal2017,
author = {Wersal, C.},
booktitle = {Ph.D. Thesis},
title = {{Neutral atom dynamics and plasma turbulence in the tokamak periphery}},
volume = {7722},
year = {2017},
number = {Lausanne},
school = {EPFL}
}

@article{Wersal2015,
author = {Wersal, C. and Ricci, P.},
doi = {10.1088/0029-5515/55/12/123014},
issn = {17414326},
journal = {Nucl. Fusion},
number = {12},
pages = {123014},
publisher = {IOP Publishing},
title = {{A first-principles self-consistent model of plasma turbulence and kinetic neutral dynamics in the tokamak scrape-off layer}},
url = {http://dx.doi.org/10.1088/0029-5515/55/12/123014},
volume = {55},
year = {2015}
}

@article{Wiesen2015,
author = {Wiesen, S. and Reiter, D. and Kotov, V. and Baelmans, M. and Dekeyser, W. and Kukushkin, A. S. and Lisgo, S. W. and Pitts, R. A. and Rozhansky, V. and Saibene, G. and Veselova, I. and Voskoboynikov, S.},
doi = {10.1016/j.jnucmat.2014.10.012},
issn = {00223115},
journal = {J. Nucl. Mater.},
pages = {480--484},
publisher = {Elsevier B.V.},
title = {{The new SOLPS-ITER code package}},
url = {http://dx.doi.org/10.1016/j.jnucmat.2014.10.012},
volume = {463},
year = {2015}
}

@article{Zhang2019,
author = {Zhang, D. R. and Chen, Y. P. and Xu, X. Q. and Xia, T. Y.},
doi = {10.1063/1.5084093},
issn = {10897674},
journal = {Phys. Plasmas},
number = {1},
title = {{Self-consistent simulation of transport and turbulence in tokamak edge plasma by coupling SOLPS-ITER and BOUT++}},
volume = {26},
year = {2019}
}

@book{Atkinson1997, 
series={Cambridge Monographs on Applied and Computational Mathematics},
title={The Numerical Solution of Integral Equations of the Second Kind}, 
publisher={Cambridge University Press}, 
author={Atkinson, Kendall E.}, 
year={1997}}

@book{Balay1998,
author = {Balay, Satish and Gropp, William and McInnes, Lois Curfman and Smith, Barry F},
number = {17},
publisher = {Argonne National Laboratory},
title = {{PETSc, the portable, extensible toolkit for scientific computation}},
volume = {2},
year = {1998}
}

@article{Oliveira2022,
arxivId = {2109.01618},
author = {Oliveira, D. S. and Body, T. and Galassi, D. and Theiler, C. and Laribi, E. and Tamain, P. and Stegmeir, A. and Giacomin, M. and Zholobenko, W. and Ricci, P. and Bufferand, H. and Boedo, J. A. and Ciraolo, G. and Colandrea, C. and Coster, D. and {De Oliveira}, H. and Fourestey, G. and Gorno, S. and Imbeaux, F. and Jenko, F. and Naulin, V. and Offeddu, N. and Reimerdes, H. and Serre, E. and Tsui, C. K. and Varini, N. and Vianello, N. and Wiesenberger, M. and W{\"{u}}thrich, C.},
doi = {10.1088/1741-4326/ac4cde},
eprint = {2109.01618},
issn = {17414326},
journal = {Nucl. Fusion},
number = {9},
publisher = {IOP Publishing},
title = {{Validation of edge turbulence codes against the TCV-X21 diverted L-mode reference case}},
volume = {62},
year = {2022}
}

@article{Dudson2009,
title = {BOUT++: A framework for parallel plasma fluid simulations},
journal = {Computer Physics Communications},
volume = {180},
number = {9},
pages = {1467-1480},
year = {2009},
issn = {0010-4655},
doi = {https://doi.org/10.1016/j.cpc.2009.03.008},
url = {https://www.sciencedirect.com/science/article/pii/S0010465509001040},
author = {B.D. Dudson and M.V. Umansky and X.Q. Xu and P.B. Snyder and H.R. Wilson}
}

@article{Tamain2016,
author = {Tamain, P. and Bufferand, H. and Ciraolo, G. and Colin, Clothilde and Galassi, Davide and Ghendrih, Ph and Schwander, Frederic and Serre, Eric},
year = {2016},
month = {05},
pages = {},
title = {The TOKAM3X code for edge turbulence fluid simulations of tokamak plasmas in versatile magnetic geometries},
volume = {321},
journal = {Journal of Computational Physics},
doi = {10.1016/j.jcp.2016.05.038}
}

@article{Zhu2018,
title = {GDB: A global 3D two-fluid model of plasma turbulence and transport in the tokamak edge},
journal = {Computer Physics Communications},
volume = {232},
pages = {46-58},
year = {2018},
issn = {0010-4655},
doi = {https://doi.org/10.1016/j.cpc.2018.06.002},
url = {https://www.sciencedirect.com/science/article/pii/S001046551830208X},
author = {Ben Zhu and Manaure Francisquez and Barrett N. Rogers}
}

@article{Stegmeir2018,
author = {Stegmeir, Andreas and Coster, David and Ross, Alexander and Maj, Omar and Lackner, Karl and Poli, Emanuele},
year = {2018},
month = {01},
pages = {},
title = {GRILLIX: a 3D turbulence code based on the flux-coordinate independent approach},
volume = {60},
journal = {Plasma Physics and Controlled Fusion},
doi = {10.1088/1361-6587/aaa373}
}

@article{Madsen2018,
title = "Plasma particle sources due to interactions with neutrals in a turbulent scrape-off layer of a toroidally confined plasma",
author = "Thrys{\o}e, {Alexander Simon} and M. L{\o}iten and J. Madsen and Volker Naulin and Nielsen, {A. H.} and Rasmussen, {J. Juul}",
year = "2018",
doi = "10.1063/1.5019662",
language = "English",
volume = "25",
journal = "Physics of Plasmas",
issn = "1070-664X",
publisher = "American Institute of Physics",
number = "3",
}

@article{Braginskii1965,
author = {Braginskii, S. I.},
journal = {Rev. Plasma Phys.},
month = {01},
pages = {205},
title = {{Transport Processes in a Plasma}},
volume = {1},
year = {1965}
}

@article{Stotler2005,
title = {Three-dimensional simulation of gas conductance measurement experiments on Alcator C-Mod},
journal = {Journal of Nuclear Materials},
volume = {337-339},
pages = {510-514},
year = {2005},
note = {PSI-16},
issn = {0022-3115},
doi = {https://doi.org/10.1016/j.jnucmat.2004.10.091},
url = {https://www.sciencedirect.com/science/article/pii/S0022311504008190},
author = {D.P. Stotler and B. LaBombard}
}

@article{Mandrekas2004,
title = {GTNEUT: A code for the calculation of neutral particle transport in plasmas based on the Transmission and Escape Probability method},
journal = {Computer Physics Communications},
volume = {161},
number = {1},
pages = {36-64},
year = {2004},
issn = {0010-4655},
doi = {https://doi.org/10.1016/j.cpc.2004.04.009},
url = {https://www.sciencedirect.com/science/article/pii/S001046550400205X},
author = {John Mandrekas}
}

@article{Rivals2025,
author = {Rivals, N. and Tamain, P. and Marandet, Y. and Bonnin, X. and Park, J.-S and Bufferand, H. and Pitts, R.A. and Falchetto, G. and Yang, H. and Ciraolo, G.},
year = {2025},
month = {01},
pages = {},
title = {SOLEDGE3X full vessel plasma boundary simulations of ITER non-active phase plasmas},
volume = {65},
journal = {Nuclear Fusion},
doi = {10.1088/1741-4326/ada2a8}
}
\end{document}